\documentclass[preprint,aps,prc,showpacs,nofootinbib,secnumarabic,floatfix]{revtex4-1}

\usepackage{graphicx}
\usepackage{bm}
\usepackage{color}
\usepackage{gensymb}
\usepackage{amsmath}
\usepackage{slashed}
\usepackage{appendix}
\usepackage{hyperref}

\newcommand{\ltsim}{\protect\raisebox{-0.5ex}{$\:\stackrel{\textstyle <}
	{\sim}\:$}}
\newcommand{\gtsim}{\protect\raisebox{-0.5ex}{$\:\stackrel{\textstyle >}
	{\sim}\:$}}

\newcommand{\bvec}[1]{\ensuremath{\boldsymbol{#1}}}

\begin{document}

\title{Cumulative proton production in $pd$ collisions}  

\author{A.B. Larionov}
\email{e-mail: larionov@theor.jinr.ru}
\affiliation{Bogoliubov Laboratory of Theoretical Physics, Joint Institute for Nuclear Research, 141980 Dubna, Russia}

\begin{abstract}
  The model of backward proton production in the exclusive reaction $pd \to ppn$ at proton beam momenta up to several GeV/c is constructed
  on the basis of Feynman diagrams for one- and two-step amplitudes. The latter include nucleon and $\Delta(1232)$ resonance intermediate scattering states
  with propagators reduced to eikonal form using the generalized eikonal approximation. The model calculations are compared with
  available experimental data on the energy spectra of protons at backward polar angles in the deuteron rest frame.
  It is shown that inclusion of two-step amplitudes significantly increases the production of backward protons with energies above $\sim$ 50 MeV, thereby significantly
  improving the agreement with experiment.
  The remaining theoretical problem related to the description of the off-shell behavior of the elementary amplitudes of the $NN \to NN$ and $NN \leftrightarrow N\Delta$
  transitions is discussed.
  Partial discrepancies between different sets of experimental data do not allow for conclusion that exotic effects, such as proton interactions with density fluctuations
  and/or $6q$ clusters in the deuteron, are present.
\end{abstract}

\maketitle

\section{Introduction}
\label{intro}

Particle production in the cumulative region, i.e., in the kinematic region forbidden for a free proton target at rest, is a long-standing problem in relativistic hadron-nucleus interactions.
Pioneering studies of the cumulative production of protons and pions in $dp$ collisions with an incident deuteron momentum of 3.3 GeV/c were performed in the 1970s
in bubble chamber experiments at JINR (see \cite{Warsaw-Dubna:1977qek} and references therein).
At high beam energies in the target nucleus rest frame,  the cumulative region is determined by the Feynman variable \cite{Frankfurt:1979sv} as
\begin{equation}
     x_F = \frac{\sqrt{m^2+\bvec{p}^2}  -p^z}{m_N} > 1~,   \label{cumulCond}
\end{equation}
where $m$ is the mass of produced particle, $m_N$ is the nucleon mass, and the axis $z$ is along the beam direction.
For proton production, the condition (\ref{cumulCond}) can be expressed as $p^z < p_\perp^2/2m_N$  or $\Theta_{\rm lab} > \arctan(2m_N/p_\perp)$,
i.e. the backward momentum hemisphere in the rest frame (r.f.) of the target nucleus is entirely included in the cumulative region.
There are several possible mechanisms of cumulative production.

One idea \cite{Burov:1976xd} is that the incoming proton interacts with a heavy object consisting of several nucleons, called a fluctuon, which is a density fluctuation
in a small volume within the nucleus \cite{Blokhintsev:1957pqs}. In particular, compact multiquark clusters can be viewed as fluctuons \cite{Burov:1983gy}.

An alternative approach to fast backward particle production based on short-range-correlated $NN$-pairs in
the nucleus was proposed in Refs.~\cite{Frankfurt:1977np,Frankfurt:1979sv}.
The interaction of an incident particle with a quasi-deuteron is described by the spectator mechanism taking into account relativistic effects in the deuteron
within the framework of the light cone formalism.
This approach describes the slopes of the momentum dependence of the proton spectra at backward angles, but still underestimates the production above 0.5 GeV/c,
leaving room for additional effects.
A similar approach, albeit in a non-relativistic framework, was developed in the work~\cite{Furui:1978tf}.

A multistep rescattering process can also potentially lead to enhanced backward production.
In the case of heavy nuclear targets, cumulative particle production was studied in the hadron cluster (fireball) model \cite{Gorenstein:1976zg,Anchishkin:1981fb}.
The influence of multiparticle interactions on backward production was studied in Ref.~\cite{Danielewicz:1990cd} based on the calculation of collision rates.
Recently, the heavy baryonic resonance and UrQMD models \cite{Motornenko:2016sfg,Panova:2019exs} have been used to describe backward production in cascade processes.
These models involve a chain of binary collision processes
$NN \to RN, RN \to RN,\ldots$, where the resonance $R$ increases its mass with each
subsequent scattering and is eventually absorbed by a nucleon $RN \to N(180\degree) N$ or decays $R \to N(180\degree) \pi$, producing a nucleon at
an angle $180\degree$ relative to the direction of the incoming proton in the r.f. of the target nucleus. In the hydrodynamic approach \cite{Dyachenko:2024nsu} cumulative production
was considered as a consequence of local thermodynamic equilibrium.  

For the $pd \to ppn$ process, the contribution of double scattering to the backward proton production was first considered in Ref.~\cite{Kopeliovich:1978cb}.
It has been shown that amplitudes with intermediate $\Delta$ resonance significantly increase the yield of backward protons above 0.3 GeV/c.
However, the authors of Ref.~\cite{Kopeliovich:1978cb} neglected the interference between one- and two-step amplitudes, which, according to Glauber theory, should be important.

In Ref.~\cite{Amelin:1978qn}, triangular graphs with pion exchange were included instead of graphs with intermediate $\Delta$ resonance, which resulted in a significant increase
of backward proton production and reasonable agreement with the data. However, the contribution of the triangular graphs with nucleon exchange was missed.

A complete account of the interference between the single and double scattering amplitudes, including the intermediate states of the nucleon and $\Delta$ resonance,
was taken in Ref.~\cite{Haneishi:1985ce}, which led to a reasonable description of the backward proton production in $pd$ collisions.
However, the authors of Ref.~\cite{Haneishi:1985ce} used a simplified assumption regarding
the amplitude $NN \to N\Delta$, considering it proportional to the amplitude  $NN \to NN$.
Moreover, they factorized the propagators of intermediate scattering states from the momentum integrals,
which is unjustified from the point of view of Glauber theory and also contradicts other authors (cf. Ref.~\cite{Sekihara:2010rw}).

A more complex calculation of multistep amplitudes for the $pd \to ppn$ process was performed in Ref.~\cite{Dakhno:1988qx}.
The double rescattering amplitudes with a fast intermediate proton were evaluated within the Glauber theory, and the double rescattering
amplitudes with a slow intermediate nucleon were evaluated using the Goldberger-Watson-Migdal correction factor.
The latter leads to the suppression of the backward proton production for momenta below $0.3$ GeV/c.
At higher momenta, the elastic and inelastic rescattering amplitudes do not lead to a significant change in the backward proton production
compared to the one-step amplitudes of the impulse approximation (IA). The calculation of Ref.~\cite{Dakhno:1988qx} significantly underestimates
the backward proton production in the process $pd \to ppn$.

Until now, most experimental studies of cumulative proton production in $pd$ collisions have concerned the inclusive reaction $pd \to p X$.
This, of course, allows for other competing channels with the contribution of pion production.
In Refs.~\cite{Braun:1978ik,Braun:1984ikd,Braun:1986xp} for proton beam momenta of $10-70$ GeV/c, the shoulder in the experimental cross section
at momenta of the proton emitted backward in the range of 0.3-0.5 GeV/c
was explained as a consequence of elastic rescattering of the pion on the spectator nucleon.
 
At the kinematic threshold (the most energetic backward proton), the residual system $X$ in the process $pd \to p X$ becomes simply a deuteron. 
In Refs.~\cite{Imambekov:1989kd,Uzikov:1998qk} the elastic process $pd \to dp$ with a proton scattered by $180\degree$ in the center-of-mass (c.m.) system was studied
in a model including one-nucleon exchange, single scattering and intermediate $\Delta$ excitation.
It is shown that the coherent sum of these three mechanisms allows one to obtain satisfactory agreement with the data on the dependence of the cross section
on the beam energy at $E_{\rm lab} < 3$ GeV.
However, the cross section at $E_{\rm lab} \simeq 1$ GeV is underestimated by approximately two times.
It was also shown that the results are very sensitive to the amplitude parameters $NN \leftrightarrow N\Delta$.

The aim of this work is to develop a model of cumulative proton production in the $pd \to ppn$ process using the generalized eikonal approximation (GEA) method,
which is an extension of the Glauber formalism and is based on Feynman graphs.
The model includes one- and two-step amplitudes with an intermediate nucleon and $\Delta$ resonance.
The calculation of propagators of intermediate scattering states
is improved compared to Refs.~\cite{Haneishi:1985ce,Dakhno:1988qx}, so that it does not include an artificial constant, as in Ref.~\cite{Haneishi:1985ce},
and does not ignore the real part, as in the Ref.~\cite{Dakhno:1988qx}.
The propagators are placed inside the momentum integrals, unlike Ref.~\cite{Haneishi:1985ce}.

In previous studies of $pd \to ppn$ reactions \cite{Frankfurt:1994nw,Frankfurt:1996uz,Larionov:2022gvn}, GEA was used to describe the kinematics with
hard $pp \to pp$ scattering, which allowed its amplitude to be factorized from integrals over the momentum transfer in two-step processes.
In contrast, the kinematics of cumulative production experiments does not allow to factorize any elementary amplitude, since the momentum transfers
at both steps can be comparable.
Another specific feature of cumulative production is that even in the IA, the cumulative proton can be not only a spectator particle,
but also one on which soft scattering occurred.
Moreover, for two-step processes, at the second step it is necessary to take into account the rescattering of both fast and slow nucleons.
The latter require special attention, since they do not necessarily move forward as assumed in most approaches based on Glauber formalism.

The structure of this work is as follows.
In sec.~\ref{model}, a model for the $pd \to ppn$ reaction amplitude is formulated starting from the one-step IA amplitudes
and then adding the two-step amplitudes with the nucleon and $\Delta$ resonance intermediate scattering states.
The propagators of the intermediate scattering states
are expressed in the eikonal form which allows to perform the integration over longitudinal component of the intermediate spectator momentum analytically.
The $NN \leftrightarrow N\Delta$ amplitudes are calculated in the pion exchange model.
The antisymmetry of the full $pd \to ppn$ reaction amplitude with respect to the interchange of the
quantum numbers of the outgoing nucleons and the isoscalar character of the deuteron wave function are explicitly taken into account.
Finally, the main uncertainty of the model arising from the off-shell kinematics of the elementary transition amplitudes is discussed.
In sec.~\ref{Xsection}, the energy differential cross sections of the backward proton production in $pd \to ppn$ process are calculated and compared with experimental data.
The contributions of the elastic and inelastic intermediate scattering states are quantified.
The discussion of the influence of the relativistic form of the deuteron vertex factor and charge exchange (CEX) processes as well of the possible future improvements
is contained in sec.~\ref{Disc}.
The summary and conclusions are given in sec.~\ref{summary}.

\section{The model}
\label{model}

\subsection{Impulse approximation amplitudes}
\label{IA}

\begin{figure}
  \begin{center}
  \includegraphics[scale = 0.4]{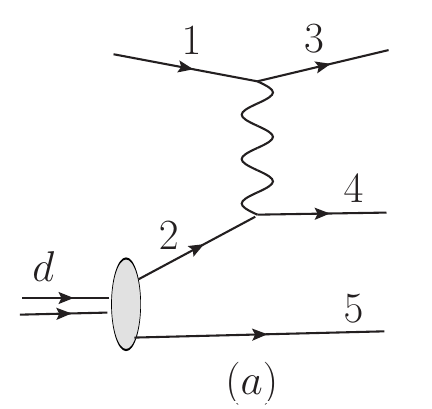}
  \includegraphics[scale = 0.4]{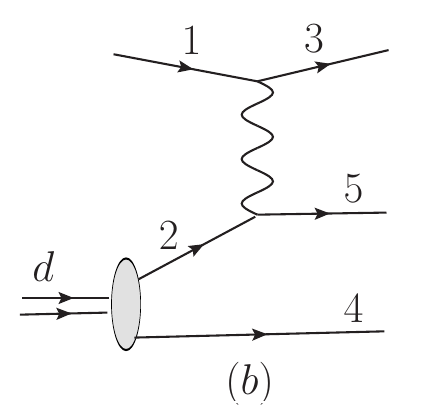}
  \includegraphics[scale = 0.4]{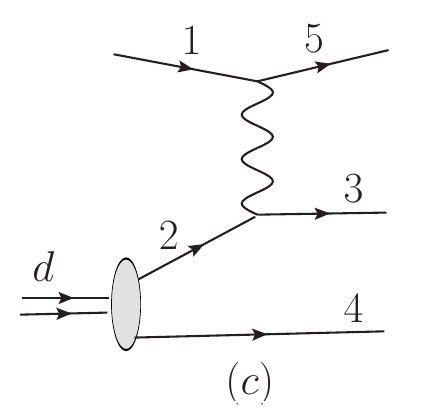}
  \includegraphics[scale = 0.4]{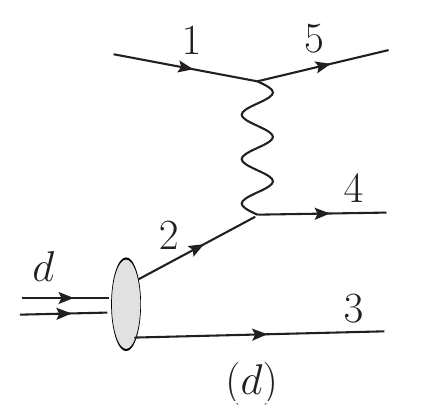}
  \end{center}
\caption{\label{fig:IA} IA diagrams for the process $p d  \to p p n$.
  The wavy lines denote elastic scattering (graphs a,b) and charge exchange (graphs c,d) amplitudes.
  d -- deuteron, 1 -- incoming proton, 2 -- struck nucleon, 3 -- outgoing fast forward proton, 4 -- outgoing slow backward proton,
  5 -- outgoing neutron.}
\end{figure}

The one-step, i.e. IA, amplitudes are shown in Fig.~\ref{fig:IA}.
The invariant amplitude (a) written for arbitrary isospin projection of the spectator nucleon 5
is expressed in the r.f. of deuteron as
\begin{equation}
  M_{\rm IA}^{(a)} = \left(\frac{2E_5 m_d}{p_2^0}\right)^{1/2} (2\pi)^{3/2} \sum_{\lambda_2,t_2} \phi_{\lambda_2\lambda_5}(\bvec{p}_2)
                [\delta_{\frac{1}{2},t_2}\delta_{-\frac{1}{2},t_5}-\delta_{\frac{1}{2},t_5}\delta_{-\frac{1}{2},t_2}] M(3,4;1,2)~,    \label{M_IA^a}
\end{equation}
where $m_d$ is the deuteron mass, $E_5=\sqrt{\bvec{p}_5^2+m_N^2}$ is the energy of the spectator nucleon, $m_N$ is the nucleon mass,
$p_2^0=m_d-E_5$ is the energy of the struck nucleon.
$\lambda_i$ and $t_i$ are, respectively, the spin and isospin projections of the nucleon $i$.  
$\phi_{\lambda_2\lambda_5}(\bvec{p}_2)$  with $\bvec{p}_2=-\bvec{p}_5$  is the spin- and momentum dependent part of the deuteron wave function (DWF) normalized as
\begin{equation}
  \sum_{\lambda_2,\lambda_5} \int d^3p |\phi_{\lambda_2\lambda_5}(\bvec{p})|^2 = 1~.    \label{DWFnorm}
\end{equation}
The prefactor in Eq.(\ref{M_IA^a}) ensures the correct flux factor in the $pd \to ppn$ cross section, see sec.~\ref{Disc} and Ref.~\cite{Larionov:2018lpk} for detail.   
The factor in square brackets  in Eq.(\ref{M_IA^a}) comes from the isospin part of the DWF having $I=0$.
$M(3,4;1,2)$ is the invariant amplitude of the elastic scattering $12 \to 34$.
The amplitudes (b),(c), and (d) are obtained from Eq.(\ref{M_IA^a}) by the permutations of quantum numbers (i.e. the momenta, and the projections of the spin and isospin)
of the outgoing nucleons 3,4, and 5, and contribute to the total amplitude with signs given by the parities of the corresponding permutations. 
Note that in Fig.~\ref{fig:IA} the physical $NN \to NN$ amplitudes are split to the $t$- and $u$-channel parts so that the partial amplitudes (a),(b),(c) and (d)
disappear for large negative four-momentum transfer squared. Thus, at high energies, the wavy lines would correspond to the pomeron or,
in the case of CEX -- to the $\pi$ and $\rho$ exchange. 
For example, one can equivalently replace $M(3,5;1,2)$ by $M_{\rm phys}(3,5;1,2) = M(3,5;1,2) - M(5,3;1,2)$
in the elastic scattering amplitude (b) and remove the CEX amplitude (c), but the separation of the $t$- and $u$-channels is better suited for the eikonal representation of the
propagator of intermediate nucleon (see sec.~\ref{El}). Since the proton 4 is flying backward, the partial amplitudes with transition $1 \to 4$
along the same continuous line are suppressed and will not be taken into account.

In the simplest approximation used in most Glauber-like calculations, the elementary elastic $NN$ scattering amplitude contains the $t$-channel part only:
\begin{equation}
  M(3,4;1,2) = 2 I_{NN} \sigma_{NN} (i+\rho_{NN}) \mbox{e}^{B_{NN}t/2}  \delta_{t_3 t_1} \delta_{t_3+t_4,t_1+t_2} \delta_{\lambda_4 \lambda_2} \delta_{\lambda_3 \lambda_1}~,    \label{M3412}
\end{equation}
where $I_{NN}=m_N\sqrt{s^2/4m_N^2-s}$ is the Moeller flux factor, $s=(p_1+p_2)^2$ and $t=(p_1-p_3)^2$ are the Mandelstam variables,
$\sigma_{NN}$ is the total $NN$ interaction cross section,
$B_{NN}$ is the slope parameter of the $t$-dependence,
and $\rho_{NN} = \mbox{Re}M(0)/\mbox{Im}M(0)$ is the ratio of the real-to-imaginary part of the forward scattering amplitude.

The physical $NN$ scattering amplitude is antisymmetric with respect to the interchange of either final or initial nucleons.
In the case of $pp$ elastic scattering, $M_{\rm phys}(3,4;1,2)=M(3,4;1,2)- M(4,3;1,2)$ where the second term corresponds to the $u$-diagram.
In the case of $pn$ elastic scattering, the $u$-diagram describes the CEX
$pn \to np$ scattering  and has to be evaluated separately. One can adopt the following form of the CEX amplitude:
\begin{equation}
  M_{\rm CEX}(4,3;1,2) = -a^{1/2} 2 I_{pn} \sigma_{pn} (1+\rho_{pn}^2)^{1/2} \mbox{e}^{B_{NN}u/2} \delta_{-t_4 t_1} \delta_{t_3+t_4,t_1+t_2}
                        \frac{\bvec{\sigma}_{\lambda_4 \lambda_1}\bvec{\sigma}_{\lambda_3 \lambda_2}}{\sqrt{3}}~,    \label{M4312_CEX}
\end{equation}
where $u=(p_1-p_4)^2$ and $\bvec{\sigma}$ is the spin Pauli matrix. The quantity $a$ determines the relative strength
of the $u$-channel term in the $pn$ elastic differential cross section \cite{Cugnon:1996kh}:
\begin{equation}
      a= \left\{
     \begin{array}{ll}
       1  & \mbox{for}~p_{\rm lab} < 0.8 \\
       0.64/p_{\rm lab}^2 & \mbox{for}~0.8 < p_{\rm lab}  
     \end{array}    
     \right.                         \label{a}
\end{equation}
with $p_{\rm lab}$ in GeV/c. The spin structure of the CEX amplitude and the appearance of the ``-'' sign in Eq.(\ref{M4312_CEX})
are in  agreement with effective potential parameterization of the DWBA $\pi$- and $\rho$-exchange model of Ref.~\cite{Kelkar:1997cr}
(see Eq.(24) of that paper).
Parameterization of Eq.(\ref{M4312_CEX}) is certainly quite rough but still enough to estimate the role of the CEX.  
Note that the physical $pn$ elastic scattering amplitude is then given by the expression $M_{\rm phys}(3,4;1,2)=M(3,4;1,2)- M_{\rm CEX}(4,3;1,2)$
which is, however, not explicitly antisymmetric. \footnote{The physical $pn$ amplitude will become antisymmetric if Eq.(\ref{M3412}) and Eq.(\ref{M4312_CEX}) are viewed as final results
of calculation based on some interaction Lagrangian. Having this in mind, the explicitly antisymmetric form of the physical amplitude valid both for $pp$ and $pn$ scattering is
$M_{\rm phys}(3,4;1,2)=M(3,4;1,2)- M_{\rm CEX}(4,3;1,2) - M(4,3;1,2) + M_{\rm CEX}(3,4;1,2)$.}

\begin{figure}
  \begin{center}
    \begin{tabular}{cc}
  \includegraphics[scale = 0.6]{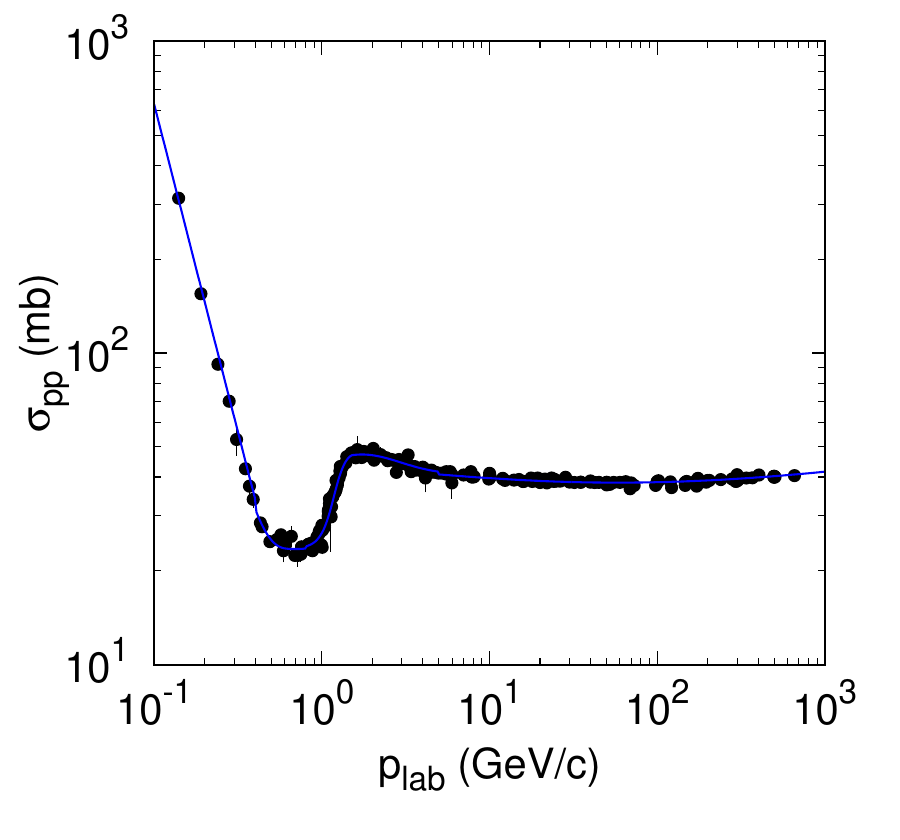} &
  \includegraphics[scale = 0.6]{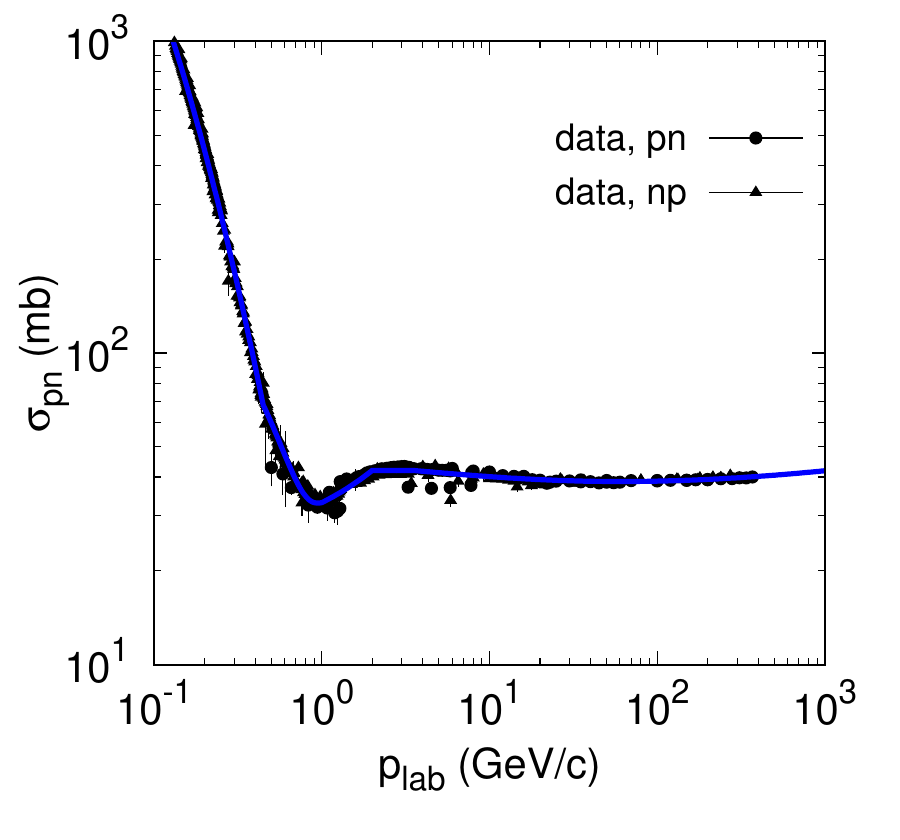} \\
  \includegraphics[scale = 0.6]{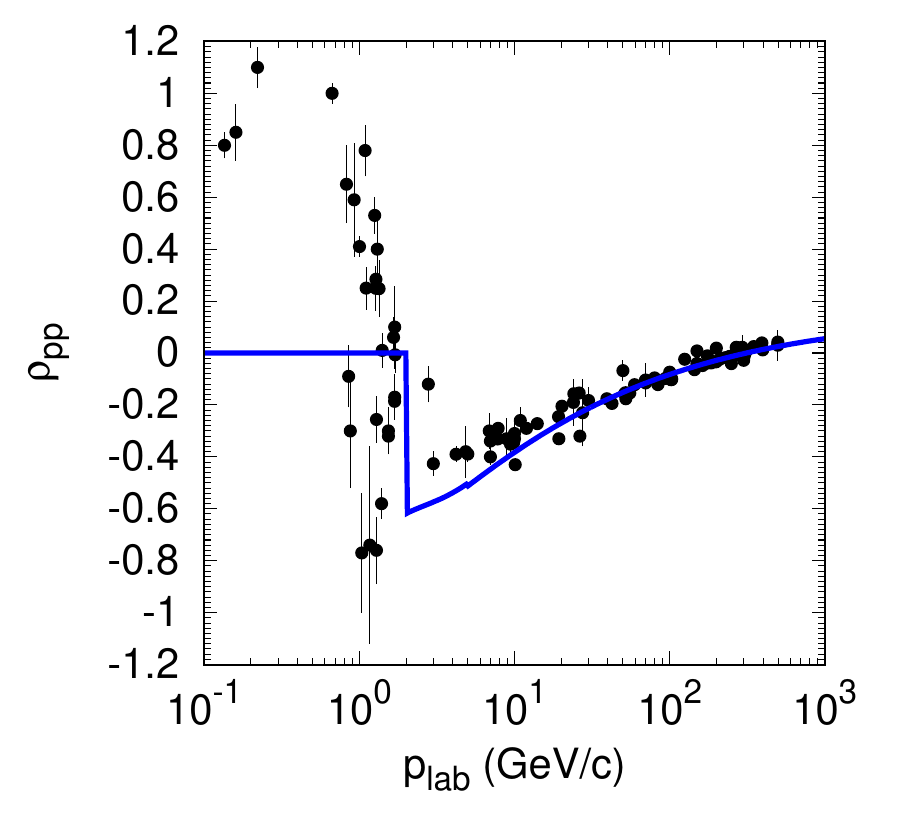} &
  \includegraphics[scale = 0.6]{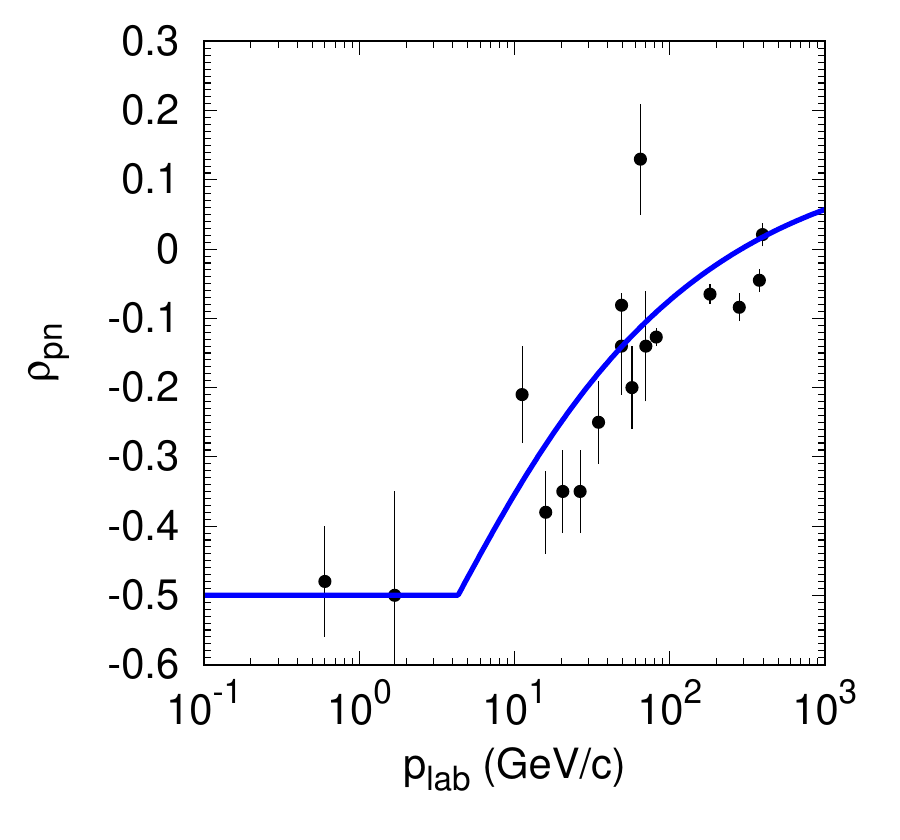} \\
    \end{tabular}
  \end{center}
  \caption{\label{fig:sigtot&alpha} Total cross section and the ratio of the real-to-imaginary part
    of the forward elastic scattering amplitude for $pp$ and $pn$ collisions vs beam momentum. Data are from Ref.~\cite{ParticleDataGroup:2022pth}.}
\end{figure}

Numerical calculations have been performed with beam-momentum dependent parameterizations of $\sigma_{pp},\sigma_{pn},\rho_{pp},$ and $\rho_{pn}$.
Thereby, the beam momentum $p_{\rm lab}$ has been calculated using the relation $p_{\rm lab}=\sqrt{s^2/4m_N^2-s}$ that assumes scattering on free nucleon.
The total cross section $\sigma_{pp} (\sigma_{pn})$  at $p_{\rm lab} < 5 (3.5)$ GeV/c is taken in the parameterization of Ref.~\cite{Cugnon:1996kh}.
The parameter $\rho_{pp} (\rho_{pn})$ at $p_{\rm lab} < 2 (4.3)$ GeV/c is set to 0 (-0.5).  
At higher $p_{\rm lab}$ values, the Regge-Gribov fits from Ref.~\cite{Patrignani:2016xqp} are used for $\sigma_{pp},\sigma_{pn},\rho_{pp},$ and $\rho_{pn}$.
Fig.~\ref{fig:sigtot&alpha} shows the used parameterizations in comparison with experimental data. The total $pp$ and $pn$ cross sections
are in perfect agreement with experiment in a full available beam momentum range. The parameters $\rho_{pp}$ and $\rho_{pn}$ are in good agreement
with data at $p_{\rm lab} \gtsim 10$ GeV/c. At lower beam momenta, the data for $pp$ strongly scatter while the data for $pn$ are extremely scarce.
Thus, setting $\rho_{pp}=0$ and $\rho_{pn}=-0.5$ seems reasonable, at least at the beam momenta above 1 GeV/c relevant for the present study.

\begin{figure}
  \begin{center}
    \includegraphics[scale = 0.6]{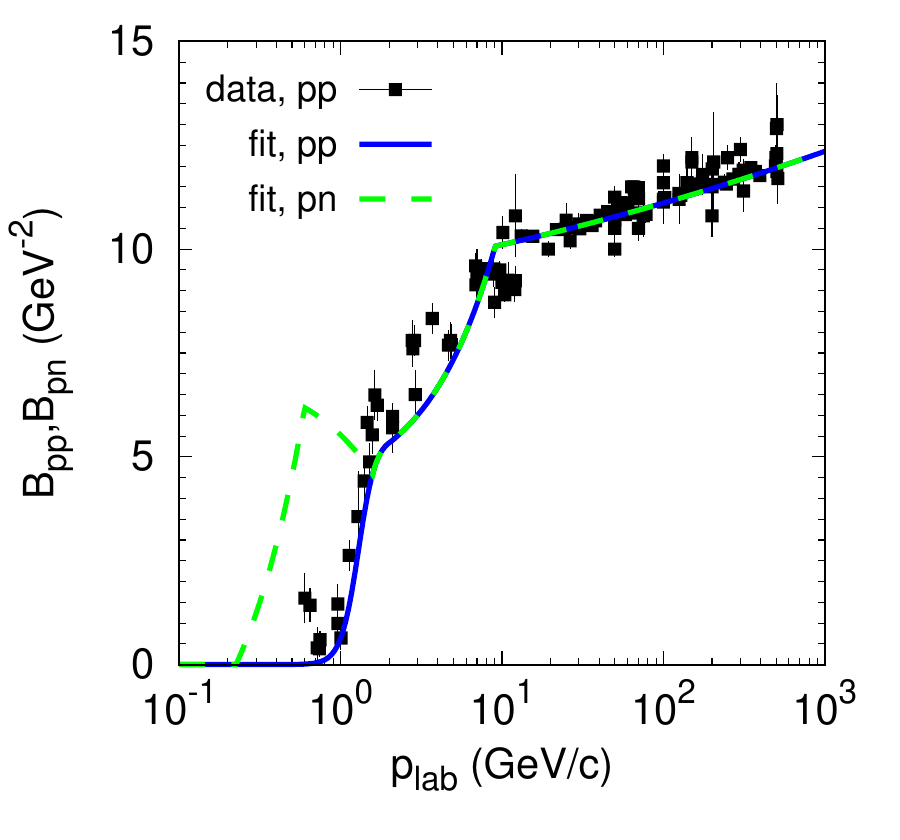}
  \end{center}
  \caption{\label{fig:B}
    Slope parameter of the $t$-dependence for the elastic scattering amplitudes $pp$ and $pn$. Data for $pp$ collisions are taken from Ref.~\cite{Okorokov:2015bha}.}
\end{figure}

The slope parameters $B_{pp}$ and $B_{pn}$ at  $p_{\rm lab} < 9.1$ GeV/c were taken in the parameterizations of Ref.~\cite{Cugnon:1996kh}.
At higher beam momenta, the ``PYTHIA'' parameterization for $B_{pp}$ from Ref.~\cite{Falter:2004uc} was applied.
Fig.~\ref{fig:B} shows the beam momentum dependence of parameters $B_{pp}$ and $B_{pn}$ in  comparison with experimental data (for $pp$ only).
The slope parameter for $pp$ elastic collisions is in a reasonable agreement with data. For the slope parameter for $pn$ elastic collisions,
the experimental data are missing, although the data on $d\sigma/dt$ at $p_{\rm lab}=1.196$ GeV/c (see Fig.~6 in Ref.~\cite{Cugnon:1996kh})
are described quite well. 

Since the full amplitude should be antisymmetric with respect to the interchange of any pair of identical fermions, both of them are either in the initial or final state,
the amplitudes (a) and (c) should contribute with ``+'' sign while the amplitudes (b) and (d) -- with ``-'' sign.
This leads to the following expression for the total IA amplitude:  
\begin{eqnarray}
  \lefteqn{M_{\rm IA} = M_{\rm IA}^{(a)} - M_{\rm IA}^{(b)} + M_{\rm IA}^{(c)} - M_{\rm IA}^{(d)} = 2 (2m_d)^{1/2} (2\pi)^{3/2}} &&   \nonumber \\
  &\times& \left\{
            \delta_{\lambda_3 \lambda_1} \left[  \left(\frac{E_5}{m_d-E_5}\right)^{1/2} I_{pp} \sigma_{pp} (i+\rho_{pp}) \mbox{e}^{B_{pp}(p_1-p_3)^2/2} \phi_{\lambda_4\lambda_5}(-\bvec{p}_5) \right. \right. \nonumber \\
  &&                               \left. + \left(\frac{E_4}{m_d-E_4}\right)^{1/2} I_{pn} \sigma_{pn} (i+\rho_{pn}) \mbox{e}^{B_{pn}(p_1-p_3)^2/2} \phi_{\lambda_5\lambda_4}(-\bvec{p}_4)  \right] \nonumber \\
  && + \frac{\bvec{\sigma}_{\lambda_5 \lambda_1}}{\sqrt{3}} \left[ \left(\frac{E_4}{m_d-E_4}\right)^{1/2} a^{1/2} I_{pn} \sigma_{pn} (1+\rho_{pn}^2)^{1/2} \mbox{e}^{B_{pn}(p_1-p_5)^2/2}
                                     \sum_{\lambda_2} \bvec{\sigma}_{\lambda_3 \lambda_2} \phi_{\lambda_2\lambda_4}(-\bvec{p}_4) \right.   \nonumber \\
  &&               \left.  \left.                           - \left(\frac{E_3}{m_d-E_3}\right)^{1/2} a^{\prime 1/2} I_{pn}^\prime \sigma_{pn}^\prime (1+\rho_{pn}^{\prime 2})^{1/2} \mbox{e}^{B_{pn}^\prime(p_1-p_5)^2/2}
                                     \sum_{\lambda_2} \bvec{\sigma}_{\lambda_4 \lambda_2} \phi_{\lambda_2\lambda_3}(-\bvec{p}_3) \right] \right\}~,      \label{M_IA_tot}
\end{eqnarray}
where the kinematics and the isospins of the nucleons are chosen as explained in the caption to Fig.~\ref{fig:IA}.
The primes in the last term of Eq.(\ref{M_IA_tot}) are introduced to distinguish $s=(p_5+p_3)^2$ in the amplitude (c) from $s^\prime = (p_5+p_4)^2$ in the amplitude (d).  
The second term in Eq.(\ref{M_IA_tot}) corresponds to the spectator mechanism of cumulative proton production. 
\footnote{A similar Eq.(1) of Ref.~\cite{Dakhno:1988qx} can be obtained from the first two terms of our Eq.(\ref{M_IA_tot}) by
setting $E_4=E_5=m_N$, i.e. neglecting the Fermi motion, replacing $I_{NN} \to s/2$, which is valid at high energies,
and including the factor $(2\pi)^{3/2}$ in the definition of DWF.}

\subsection{Elastic rescattering amplitudes}
\label{El}

\begin{figure}
  \begin{center}
  \includegraphics[scale = 0.4]{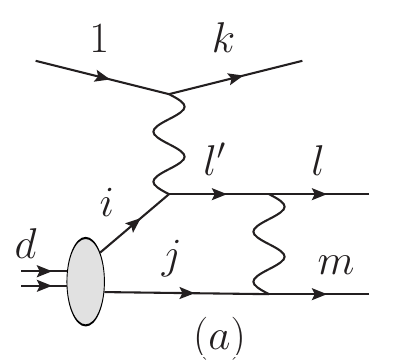}
  \includegraphics[scale = 0.4]{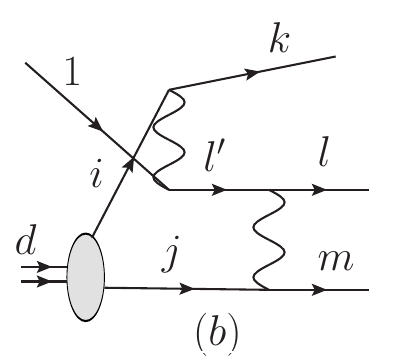}
  \end{center}
  \caption{\label{fig:el} Elastic rescattering and CEX diagrams for the process $p d \to p p n$.
    The pair $i,j$ represents permutations of nucleons 2 and $5^\prime$, and the triple $k,l,m$ represents permutations of nucleons 3,4, and 5.
    The particles are denoted as follows: d -- deuteron, 1 -- incoming proton, 2 -- proton of the deuteron, $5^\prime$ -- neutron of the deuteron, 3 -- outgoing forward proton,
    4 -- outgoing backward proton, 5 -- outgoing neutron.
    $l^\prime$ denotes the intermediate nucleon scattering state.
    Allowed are six graphs of type (a) with $k \neq 4$ and seven graphs of type (b) with $l \neq 4$.}
\end{figure}

The next step is to include rescattering corrections.
Possible two-step amplitudes with an intermediate nucleon are shown in Fig.~\ref{fig:el}.
The total elastic rescattering amplitude $M_{\rm el}$ is given by the antisymmetrized sum of six amplitudes of type (a) and seven amplitudes of type (b)
allowed by the kinematics and nucleon charge conservation in elementary $NN \to NN$ transitions. 
By keeping only the contribution of the particle pole of the propagator of intermediate spectator $j$ leads to the following expression 
\footnote{From now on, for brevity, the spin indices of the DWF and summations over
them are suppressed, where this does not lead to misunderstanding.}:
\begin{eqnarray}
  M_{\rm el} &=& -\sum_{k,l,m \in \{3,4,5\}} (-1)^{\eta_{k,l,m}} \sum_{i,j \in \{2,5^\prime\}} (-1)^{\eta_{i,j}}  \int \frac{d^3p_j}{(2\pi)^{3/2}} 
            \left(\frac{m_d}{2E_j(m_d-E_j)}\right)^{1/2} \phi(-\bvec{p}_j) \nonumber \\
           && \times \frac{M(l,m;l^\prime,j)}{p_{l^\prime}^2-m_N^2+i0} [ M(k,l^\prime;1,i) (1-\delta_{k4})
                                                                  - M(k,l^\prime;i,1) (1-\delta_{l4}) ] \delta_{t_k+t_{l^\prime},t_1+t_i}~,  \label{M_el}
\end{eqnarray}
where the intermediate nucleon $j$ is put on the mass shell, i.e. $p_j^0=E_j \equiv \sqrt{\bvec{p}_j^2+m_N^2}$.
In Eq.(\ref{M_el}), the signs of the partial amplitudes with spectator neutron and spectator proton are opposite
due to the antisymmetry of the isospin part of the DWF which is taken into account via the factor $(-1)^{\eta_{i,j}}$.
Another factor, $(-1)^{\eta_{k,l,m}}$, accounts for the antisymmetry of the total amplitude with respect to the interchange
of the quantum numbers of the outgoing nucleons. Here, $\eta_{i,j}$ and $\eta_{k,l,m}$ are the parities of the permutations defined such that
$\eta_{2,5^\prime}=\eta_{3,4,5}=0$. In the r.h.s. of Eq.(\ref{M_el}), the elementary amplitude $M(c,d;a,b)$ denotes either elastic scattering amplitude
of Eq.(\ref{M3412}) if $t_c=t_a$  or the CEX amplitude of Eq.(\ref{M4312_CEX}) if $t_c=-t_a$.

When calculating the elastic amplitude (\ref{M_el}) it is necessary to carefully take into account the propagator of the intermediate nucleon $l^\prime$.
The simplest and most commonly used method is to retain only the imaginary part of the propagator by replacing
$(p_{l^\prime}^2-m_N^2+i0)^{-1}$ by $-i\pi\delta(p_{l^\prime}^2-m_N^2)$, which corresponds to the simplified Glauber approximation.
\footnote{Being expressed in coordinate space  the simplification  means replacing the $\Theta$-function of the relative position of the proton and neutron
in the deuteron along the path of propagation of the fast particle, which determines the scattering order, by a factor of 1/2.}
This study deals, however, not only with intermediate fast nucleons (amplitudes of type (b)) but also with slow ones (amplitudes of type (a)).
Thus, the propagator should be treated more precisely keeping both the real and imaginary parts. To this end, the GEA method is applied.

In the GEA, the inverse propagator in Eq.(\ref{M_el}) can be written as 
\begin{equation}
  p_{l^\prime}^2-m_N^2+i0 = (p_l+q^\prime)^2-m_N^2+i0 = 2p_lq^\prime + q^{\prime\, 2} + i0 = 2|\bvec{p}_l|(-q^{\prime \tilde z} + \Delta_l + i0)~, \label{lprimeInvProp}
\end{equation}
where $q^\prime=p_m-p_j$ is the four-momentum transfer to the intermediate spectator nucleon $j$, the axis $\tilde z$ is directed along $\bvec{p}_l$,
and
\begin{equation}
  \Delta_l \equiv \frac{E_lq^{\prime 0}}{|\bvec{p}_l|} + \frac{q^{\prime\, 2}}{2|\bvec{p}_l|}
   \simeq \frac{(E_l-m_N)(E_m-m_N)}{|\bvec{p}_l|}~,    \label{Delta_l}
\end{equation}
where in the second approximate step the Fermi motion in the deuteron was neglected, $\bvec{p}_j \to 0$,
since large values of $|\bvec{p}_j|$ are suppressed by the DWF.

Using the Paris potential model \cite{Lacombe:1981eg}, the DWF can be expressed as  
\begin{equation}
     \phi(\bvec{p})=\frac{1}{\sqrt{4\pi}} \sum_{n=1}^{13} \frac{\phi_n^{\lambda_d}(\bvec{p})}{p^2+m_n^2}~,      \label{phi_decomp}
\end{equation}
where
\begin{equation}
    \phi_n^{\lambda_d}(\bvec{p}) = \left(\frac{2}{\pi}\right)^{1/2} \left(c_n+\frac{d_n}{\sqrt{8}}S(\bvec{p})\right) \chi^{\lambda_d}~,     \label{phi_j^lambda_d}
\end{equation}
with the spin tensor operator
\begin{equation}
   S(\bvec{p})=\frac{3(\bvec{\sigma}_p\bvec{p})(\bvec{\sigma}_n\bvec{p})}{p^2} 
               -\bvec{\sigma}_p\bvec{\sigma}_n~,   \label{Sspin}
\end{equation}
and $\chi^{\lambda_d}$ being the spin wave function of the $S=1$ $pn$ state with spin projection $\lambda_d=0,\pm1$.

The integral over three-momentum of the intermediate spectator neutron in Eq.(\ref{M_el}) can be taken in the rotated coordinate system with $\tilde z$ axis
along  $\bvec{p}_l$ by replacing $d^3p_j \to d^2p_{j t} dp_j^{\tilde z}$.
Since the $NN$ elastic and CEX amplitudes and the factor $\sqrt{m_d/2E_j(m_d-E_j)}$ vary slowly with momentum of the spectator neutron $\bvec{p}_j$,
it is possible to neglect their dependence on the longitudinal component $p_j^{\tilde z}$. The reason for this is the following (see also subsec.~\ref{restr} for detailed discussion
of the off-shell $NN$ amplitudes): 
In the case of Fig.~\ref{fig:el}b, the both
$NN \to NN$ amplitudes depend  on the corresponding transverse components of momentum transfer
relative to the direction of momentum $\bvec{p}_l$, since large difference $|\bvec{p}_1-\bvec{p}_l|$ is suppressed by exponential factors in Eqs.(\ref{M3412}),(\ref{M4312_CEX}).
In the case of Fig.~\ref{fig:el}a, the right $NN \to NN$ amplitude also depends on the transverse component of momentum transfer relative to the direction of $\bvec{p}_l$ 
while the left $NN \to NN$ amplitude is determined by $(p_1-p_k)^2$ and, thus, can be factorized out of momentum integral. 

Then, using the DWF representation of Eq.(\ref{phi_decomp}) and following Appendix A of Ref.~\cite{Frankfurt:1996uz}, the integration over $dp_j^{\tilde z}$
can be done analytically:
\begin{equation}
  \int \frac{dp_j^{\tilde z}}{2\pi} \frac{1}{(p_j^{\tilde z} - p_m^{\tilde z}  + \Delta_l + i0)[(p_j^{\tilde z})^2 + m_{nt}^2]}
    = \frac{1}{2m_{nt}(im_{nt} - p_m^{\tilde z}  + \Delta_l)}~,     \label{ContInt}
\end{equation}
where $m_{nt} \equiv \sqrt{p_{j t}^2+m_n^2}$ and the integral is taken by closing the integration contour in the upper part
  of $p_j^{\tilde z}$ complex plane where the only pole is at $p_j^{\tilde z}=im_{nt}$.
As a result, the elastic rescattering amplitude of Eq.(\ref{M_el}) is expressed as 
\begin{eqnarray}
  M_{\rm el} &=& -\frac{1}{8\sqrt{2} \pi} \sum_{k,l,m \in \{3,4,5\}} (-1)^{\eta_{k,l,m}} |\bvec{p}_l|^{-1} \sum_{i,j \in \{2,5^\prime\}} (-1)^{\eta_{i,j}}
                 \int d^2q_t^\prime \left(\frac{m_d}{2E_j(m_d-E_j)}\right)^{1/2} \sum_{n=1}^{13} \phi_n^{\lambda_d}(-\bvec{p}_j)  \nonumber \\
                 && \times   \frac{M(l,m;l^\prime,j) [ M(k,l^\prime;1,i) (1-\delta_{k4})
                                                    - M(k,l^\prime;i,1) (1-\delta_{l4}) ] \delta_{t_k+t_{l^\prime},t_1+t_i}}{m_{nt}(im_{nt} - p_m^{\tilde z}  + \Delta_l)}~,  \label{M_el_fin}
\end{eqnarray}                                                                  
where the integration is done over transverse to the axis $\tilde z$ component of the momentum
transfer to the intermediate spectator, $\bvec{q}_t^\prime = \bvec{p}_{m t} - \bvec{p}_{j t}$.
The three-momentum $\bvec{p}_j$ in the argument
of the DWF is obtained by expressing the three-momentum $(\bvec{p}_{j t},im_{nt})$ in the original coordinate frame
(with $z$ axis along $\bvec{p}_1$), where the deuteron spin quantization is performed.

\subsection{Inelastic rescattering amplitudes} 
\label{Inel}

The present study is limited to the resonant excitation of $\Delta(1232)$ in intermediate states and is based on the pion exchange model from Ref.~\cite{Dmitriev:1986st},
which describes well the $pp \to n\Delta^{++}$ cross section at beam momenta up to several tens of GeV/c.
The $\pi NN$ and $\pi N \Delta$ vertices are described by the following Lagrangians:
\begin{eqnarray}
  {\cal L}_{\pi NN} &=& \frac{f_{\pi NN}}{m_\pi} \bar\psi \gamma^\mu \gamma^5 \bvec{\tau} \psi \partial_\mu \bvec{\pi}~,  \label{L_piNN}\\
  {\cal L}_{\pi N \Delta} &=& \frac{f_{\pi N \Delta}}{m_\pi} \bar\psi^\mu \bvec{T} \psi \partial_\mu \bvec{\pi} + h.c.~,    \label{L_piND}
\end{eqnarray}
where $f_{\pi NN}=1.008$, $f_{\pi N \Delta}=2.202$, and $m_\pi$ is the pion mass.
$\bvec{T}$ is the isospin transition $1/2 \to 3/2$ operator (see Ref.~\cite{Brown:1975di}):
\begin{equation}
  \bvec{T}_{t_\Delta t_N} = \sum_{l=0,\pm1} <\frac{3}{2} t_\Delta|\frac{1}{2}t_N;1l> \bvec{t}^{(l)*}~, \label{T}
\end{equation}
where $\bvec{t}^{(0)}=(0,0,1), \bvec{t}^{(\pm1)}=\mp\frac{1}{\sqrt{2}}(1,\pm i,0)$ are the eigenvectors of $\hat I^2$ and $\hat I_3$
operators for $I=1$ in the Cartesian basis.
\begin{figure}
  \begin{center}
    \includegraphics[scale = 0.6]{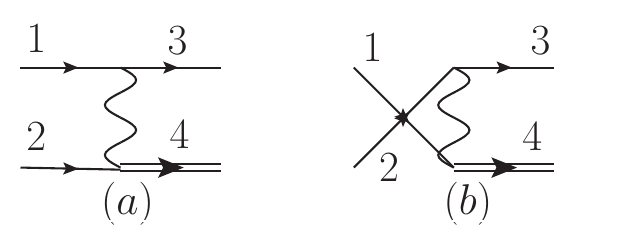}
  \end{center}
  \caption{\label{fig:NN2ND} Diagrams describing the process $N_1 N_2 \to N_3 \Delta_4$. Wavy lines represent pion exchange.}
\end{figure}
As a test, the cross section for the production of $\Delta$ in $NN$ collisions was calculated. Two partial amplitudes of this process are shown in Fig.~\ref{fig:NN2ND}.
Amplitude (a) is expressed as
\begin{eqnarray}
  M_\Delta(3,4;1,2) &=& -\frac{f_{\pi NN} f_{\pi N \Delta}F^2(t)}{m_\pi^2}
  \frac{\bar u(p_3,\lambda_3) \slashed{q} \gamma^5 u(p_1,\lambda_1) \, \bar u^\mu(p_4,\lambda_4) u(p_2,\lambda_2) q_\mu}%
       {t-m_\pi^2+i0} I_{iso}~,    \label{M^a}
\end{eqnarray}
where $q=p_1-p_3$, $t=q^2$. The isospin factor is
\begin{equation}
    I_{iso} = \bvec{T}_{t_4 t_2} \bvec{\tau}_{t_3 t_1}~.  \label{Iiso}
\end{equation}
A simple calculation gives $I_{iso}=\pm \sqrt{2}$ for $|t_4|=3/2$ and $I_{iso}=\pm \sqrt{2/3}$ for $|t_4|=1/2$
where the upper (lower) sign corresponds to $t_3=+1/2 (-1/2)$. 
According to Ref.~\cite{Dmitriev:1986st}, the $\pi NN$ and $\pi N\Delta$ vertex form factors are chosen as follows:
\begin{equation}
  F(t) = \frac{\Lambda^2-m_\pi^2}{\Lambda^2-t}~,     \label{FF}
\end{equation}
with a cutoff parameter $\Lambda=0.63$ GeV.
The Dirac spinors of the nucleons are normalized as $\bar u(p,\lambda) u(p,\lambda) = 2m_N$
while the Rarita-Schwinger vector-spinors of the $\Delta$ resonance -- as $\bar u^\mu(p_\Delta,\lambda_\Delta) u_\mu(p_\Delta,\lambda_\Delta) = -2m_\Delta$,
where the $\Delta$ is assumed to be produced on the mass shell, i.e. $p_\Delta^2=m_\Delta^2$, $m_\Delta=1.232$ GeV.  The amplitude (b) is obtained from Eq.(\ref{M^a})
by replacing $1 \leftrightarrow 2$ and $t \to u=(p_3-p_2)^2$. The angular differential cross section is expressed as
\begin{equation}
  \frac{d\sigma}{d\Theta_{c.m.}} = \frac{q_{f} \sum\limits_{\lambda_1,\lambda_2,\lambda_3,\lambda_4}|M_\Delta(3,4;1,2)-M_\Delta(3,4;2,1)|^2}{128\pi q_{i} s} \sin\Theta_{c.m.}~, \label{dsigmadThetacm}
\end{equation}
where $\Theta_{c.m.}$ is the polar scattering angle in the c.m. frame, $s=(p_1+p_2)^2$,
$q_{i}=(s/4-m_N^2)^{1/2}$ and $q_{f}=[(s-m_\Delta^2+m_N^2)^2/4s-m_N^2]^{1/2}$ are the initial and final c.m. momenta, respectively.
In Eq.(\ref{dsigmadThetacm}) the spreading width of the $\Delta$ resonance is for simplicity neglected which limits the applicability
of this formula by collisions well above pion production threshold, i.e. at $\sqrt{s} \gg 2m_N+m_\pi$.
\begin{figure}
  \begin{center}
    \includegraphics[scale = 0.6]{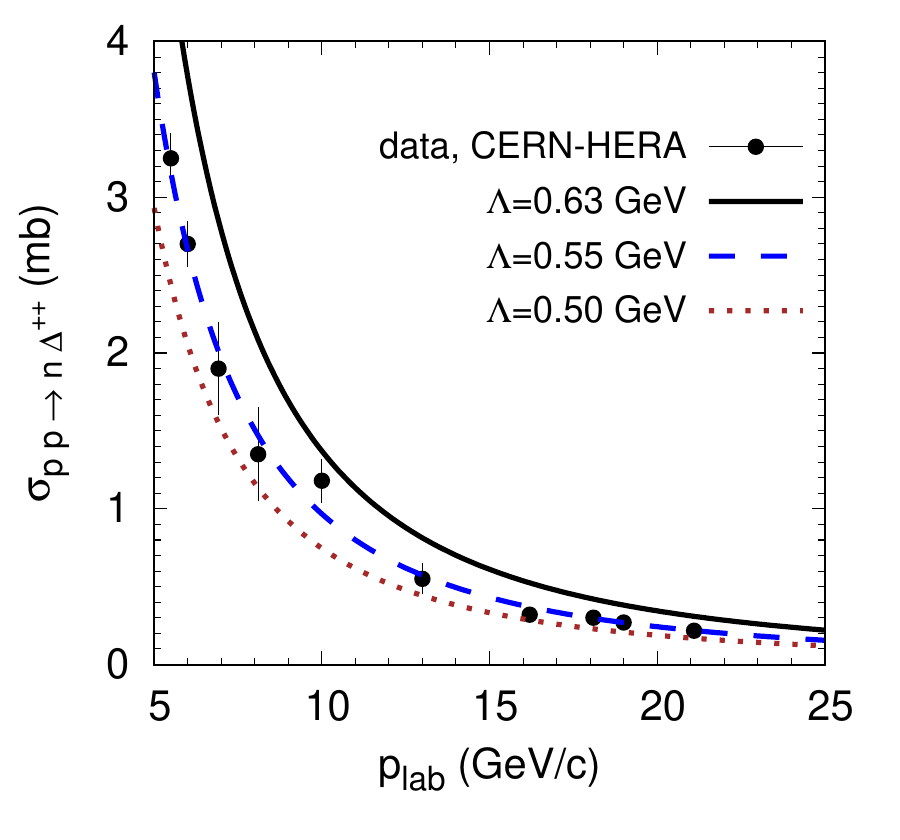}
  \end{center}
  \caption{\label{fig:sig_pp2nd++} Integrated cross section of $pp \to n\Delta^{++}$ vs proton beam momentum.
           Data are from Ref.~\cite{Flaminio:1984gr}.}
\end{figure}
Fig.~\ref{fig:sig_pp2nd++} shows the cross section of the $pp \to n\Delta^{++}$ process, obtained by integrating Eq.(\ref{dsigmadThetacm})
over $\Theta_{c.m.}$.
The shown beam momentum range corresponds to $\sqrt{s}=3.4-7.0$ GeV, where the one-pion production cross section is dominated by the excitation
of an intermediate $\Delta$ resonance near the pole mass.
The choice of the cutoff parameter $\Lambda=0.55$ GeV agrees with the data in the best way.
However, for further numerical calculations, following Ref.~\cite{Dmitriev:1986st}, the value $\Lambda=0.63$ GeV was chosen, which corresponds to the data at lower beam momenta,
where the spectral function of the $\Delta$ resonance should be included in the calculation of the $pp \to n\Delta^{++}$ cross section.

\begin{figure}
  \begin{center}
  \includegraphics[scale = 0.4]{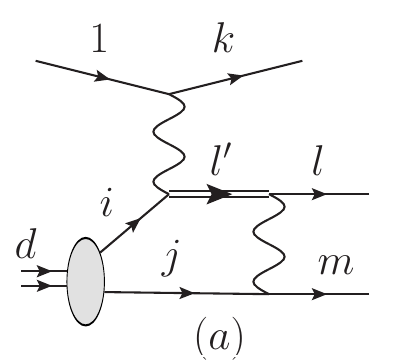}
  \includegraphics[scale = 0.4]{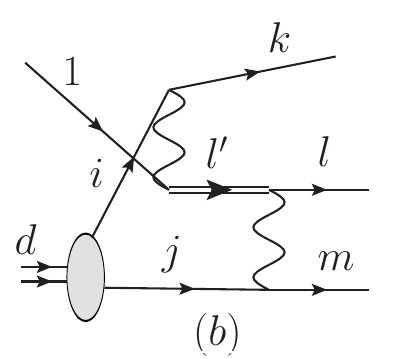}
  \end{center}
  \caption{\label{fig:inel} Diagrams with intermediate $\Delta(1232)$ resonance  for the process $p d  \to p p n$.
    Wavy lines denote pion exchange. The indices $i,j$ are, respectively, the two permutations of nucleons 2 and $5^\prime$ while $k,l,m$ are, respectively,
    the four permutations of nucleons 3,4, and 5 with $k \neq 4$ in the diagram (a) and $l \neq 4$ in the diagram (b).
    d -- deuteron, 1 -- incoming proton, 2 -- proton of the deuteron, $5^\prime$ -- neutron of the deuteron, 3 -- outgoing forward proton,
    4 -- outgoing backward proton, 5 -- outgoing neutron.
    $l^\prime$ denotes the intermediate $\Delta$ resonance.}
\end{figure}

Diagrams with intermediate $\Delta(1232)$ are shown in Fig.~\ref{fig:inel}.
Due to the possibility of exchange of charged pions, sixteen diagrams are possible: eight with slow intermediate $\Delta$ (a) and eight
with fast intermediate $\Delta$ (b).  

The antisymmetrized sum of the amplitudes of Fig.~\ref{fig:inel} is expressed as
\begin{eqnarray}
  M_{\rm inel} &=&   \sum_{i,j \in \{2,5^\prime\}}~~\sum_{k,l,m \in \{3,4,5\}} (-1)^{\eta_{i,j}} (-1)^{\eta_{k,l,m}} 
              \int \frac{d^3p_j}{(2\pi)^{3/2}} \left(\frac{m_d}{2E_j(m_d-E_j)}\right)^{1/2} \phi(-\bvec{p}_j) \nonumber \\
             && \times [(1-\delta_{k4})M(k,l,m;1,i,j) -(1-\delta_{l4}) M(k,l,m;i,1,j) ]~,    \label{M_inel}
\end{eqnarray}
where the nucleon $j$ is set on the mass shell.
$M(k,l,m;1,i,j)$ and $M(k,l,m;i,1,j)$ are the $3 \to 3$ transition amplitudes obtained by removing the deuteron decay vertices
in graphs (a) and (b), respectively. 
As in the case of the elastic rescattering amplitude (Eq.(\ref{M_el})), the factors $(-1)^{\eta_{i,j}}$ and $(-1)^{\eta_{k,l,m}}$
in Eq.(\ref{M_inel}) take into account, respectively, the antisymmetry of the isospin part of the DWF and the antisymmetry of the total amplitude
with respect to the interchange of quantum numbers of the outgoing nucleons.

The invariant amplitudes of the $3 \to 3$ transitions appearing in Eq.(\ref{M_inel})
can be calculated in fourth-order perturbation theory using the standard Feynman rules and the interaction Lagrangians (\ref{L_piNN}),(\ref{L_piND}),
which leads to the following expression:
\begin{eqnarray}
  M(k,l,m;1,i,j) &=& -\left(\frac{f_{\pi NN} f_{\pi N \Delta} F(q^2) F(q^{\prime\,2})}{m_\pi^2}\right)^2
  \frac{\bar u(p_k,\lambda_k) \slashed{q} \gamma^5 u(p_1,\lambda_1) \, \bar u(p_m,\lambda_m) \slashed{q}^\prime \gamma^5 u(p_j,\lambda_j)}%
       {(q^2-m_\pi^2+i0)(q^{\prime\,2}-m_\pi^2+i0)} \nonumber \\
      && \times \bar u(p_l,\lambda_l) G^{\mu\nu}(p_{l^\prime}) q_\mu^\prime q_\nu u(p_i,\lambda_i) \, {\cal I}_{\rm iso}~,    \label{M_klm_1ij}
\end{eqnarray}
where $q=p_1-p_k$, $q^\prime = p_m-p_j$.
The Rarita-Schwinger propagator of the $\Delta$ resonance is
\begin{equation}
  G^{\mu\nu}(p_{l^\prime}) = -\frac{(\slashed{p}_{l^\prime}+m_\Delta)}{p_{l^\prime}^2-m_{\Delta}^2+i\Gamma_\Delta m_\Delta} {\cal P}^{\mu\nu}(p_{l^\prime})~,   \label{iG^munu}
\end{equation}
where
\begin{equation}
  {\cal P}^{\mu\nu}(p_{l^\prime}) = g^{\mu\nu} - \frac{\gamma^\mu\gamma^\nu}{3}
  - \frac{2 p_{l^\prime}^\mu p_{l^\prime}^\nu}{3m_\Delta^2} + \frac{p_{l^\prime}^\mu\gamma^\nu-p_{l^\prime}^\nu\gamma^\mu}{3m_\Delta}~,   \label{calP^munu}
\end{equation}
with $\Gamma_\Delta=0.118$ GeV being the width of the $\Delta$ resonance. 
For arbitrary isospin states of the incoming and outgoing nucleons, the isospin factor is defined as
\begin{equation}
  {\cal I}_{\rm iso} = (\bvec{T}^\dag_{t_l t_{l^\prime}} \cdot \bvec{\tau}_{t_m t_j}) (\bvec{\tau}_{t_k t_1} \cdot  \bvec{T}_{t_{l^\prime} t_i})~.   \label{calI_iso}
\end{equation}
After somewhat lengthy but straightforward calculation the following result can be obtained for transitions satisfying
the necessary isospin conservation ($t_1+t_i+t_j = t_k+t_l+t_m$): ${\cal I}_{\rm iso}=\pm2$ for $|t_{l^\prime}|=3/2$,
${\cal I}_{\rm iso}=\pm2/3$ for $|t_{l^\prime}|=1/2$. Here, the upper (lower) sign corresponds to the case when $t_k=t_j$ ($t_k=-t_j$).

The $\Delta$-resonance propagator can be transformed using the GEA, similar to the elastic rescattering case.
For this purpose, it is convenient to define a new quantity, which for brevity will be called the "reduced amplitude":
\begin{equation}
   M_{\rm red}(k,l,m;1,i,j) \equiv  M(k,l,m;1,i,j)\, (p_{l^\prime}^2-m_{\Delta}^2+i\Gamma_\Delta m_\Delta)~.      \label{M_klm_1ij_red}  
\end{equation}
If one considers the $\Delta$ resonance as a stable particle that can appear in the final or initial state of the $2 \to 2$ transition,
then the reduced amplitude will simply be the product of the invariant amplitudes of $\Delta$ production and absorption,
summed over the spin projection of the $\Delta$:
\begin{equation}
  M_{\rm red}(k,l,m;1,i,j) = - \sum_{\lambda_{l^\prime}} M_\Delta^*(l^\prime,j;l,m) M_\Delta(k,l^\prime;1,i)~,
\end{equation}
where $M_\Delta(k,l^\prime;1,i)$ is the invariant amplitude of the process $1 i \to k l^\prime$, see Eq.(\ref{M^a}).

Similar to Eq.(\ref{lprimeInvProp}), the inverse propagator of the $\Delta$ resonance can be rewritten in the eikonal form: 
\begin{eqnarray}
  && p_{l^\prime}^2-m_\Delta^2+i\Gamma_\Delta m_\Delta= (p_l+q^\prime)^2-m_\Delta^2+i\Gamma_\Delta m_\Delta
  = m_N^2 + 2p_lq^\prime + q^{\prime\, 2} -m_\Delta^2+i\Gamma_\Delta m_\Delta \nonumber \\
  && = 2|\bvec{p}_l|(-q^{\prime \tilde z} + \Delta_l +  i{\cal G}_l)~, \label{lprimeInvProp_Delta}
\end{eqnarray}
with  
\begin{eqnarray}
  \Delta_l &\equiv& \frac{E_lq^{\prime 0}}{|\bvec{p}_l|} + \frac{q^{\prime\, 2}+m_N^2-m_\Delta^2}{2|\bvec{p}_l|}
  \simeq \frac{(E_l-m_N)(E_m-m_N)}{|\bvec{p}_l|} + \frac{m_N^2-m_\Delta^2}{2|\bvec{p}_l|}~,    \label{Delta_l_Delta} \\
  {\cal G}_l &\equiv&  \frac{\Gamma_\Delta m_\Delta}{2|\bvec{p}_l|}~.   \label{calg_l}
\end{eqnarray}
In the second step of Eq.(\ref{Delta_l_Delta}), the Fermi motion in the deuteron was neglected.

Following the line of derivation of Eq.(\ref{M_el_fin}) in subsec. \ref{El},
one can integrate over the longitudinal component of the intermediate spectator momentum in Eq.(\ref{M_inel})
which yields
\begin{eqnarray}
  M_{\rm inel} &=& \frac{1}{8\sqrt{2} \pi} \sum_{k,l,m \in \{3,4,5\}} (-1)^{\eta_{k,l,m}}|\bvec{p}_l|^{-1}~~ \sum_{i,j \in \{2,5^\prime\}} (-1)^{\eta_{i,j}} 
                 \int d^2q_t^\prime  \left(\frac{m_d}{2E_j(m_d-E_j)}\right)^{1/2} \sum_{n=1}^{13}  \nonumber \\
             && \times \phi_n^{\lambda_d}(-\bvec{p}_j)
                \frac{(1-\delta_{k4})M_{\rm red}(k,l,m;1,i,j)-(1-\delta_{l4}) M_{\rm red}(k,l,m;i,1,j)}{m_{nt}(im_{nt} - p_m^{\tilde z}  + \Delta_l + i{\cal G}_l)}~. \label{M_inel_fin}
\end{eqnarray}
The notations and definitions in Eq.(\ref{M_inel_fin}) are quite similar to those already introduced in subsec. \ref{El},
but they are given here again for completeness.
The axis $\tilde z$ is directed along three-momentum of the outgoing nucleon $l$ in the middle line of the graphs in Fig.~\ref{fig:inel}.
The integration is done over transverse with respect to $\tilde z$ component of the momentum transfer to the intermediate spectator
nucleon $j$ which is put on the mass shell.
The transverse mass is defined as $m_{nt} \equiv \sqrt{\bvec{p}_{j t}+m_n^2}$. The components of the three-momentum $\bvec{p}_j$ in the argument of the DWF
are obtained by expressing the three-vector $(\bvec{p}_{j t},im_{nt})$ in the original coordinate frame.

\subsection{Off-shell amplitudes and integration restrictions}
\label{restr}

When deriving the final expressions for the elastic and inelastic partial rescattering amplitudes, it was assumed that 
the elementary amplitudes in the integrals over the three-momentum $\bvec{p}_j$ of the intermediate spectator, Eqs.(\ref{M_el}),(\ref{M_inel}),
do not depend on the longitudinal component $p_j^{\tilde z}$.
\footnote{The reason for this assumption was that the elementary amplitudes change only
slowly with $|\bvec{p}_j|$, and large values of $|\bvec{p}_j|$ are suppressed by the DWF. 
In contrast, the propagator of the intermediate state
$l^\prime$ has a pole with respect to $p_j^{\tilde z}$
and therefore can not be factorized out of the integral over $p_j^{\tilde z}$, Eq.(\ref{ContInt}).}
However, in practical calculations it is anyway necessary to specify the kinematics by which the elementary amplitudes are determined.

In line with the Glauber theory, the elementary transition amplitudes should be evaluated at the on-shell kinematics
which is determined by the pole of the propagator of the intermediate state $l^\prime$:
\begin{equation}
  q^{\prime \tilde z} = \Delta_l~.    \label{onsCond}
\end{equation}
Given Eq.(\ref{onsCond}), the four-momenta of all intermediate states are completely determined
from the four-momentum conservation at the vertices.
However, the condition (\ref{onsCond}) does not completely eliminate the problems associated with off-shell particles. Since the Dirac spinor can only be defined for
a timelike fermion, the integration region was restricted to the conditions that the struck nucleon $i$ has positive energy and its four-momentum is timelike.
Moreover, due to the approximations in the calculation of $\Delta_l$'s (Eqs.(\ref{Delta_l}),(\ref{Delta_l_Delta})),
the graphs (b) of Figs.~\ref{fig:el},\ref{fig:inel} with a fast outgoing nucleon $l$ in the middle line can still give the kinematics with a timelike
first exchange four-momentum $q$ (left wavy line).
Thus, the constraint $q^2 < 0$ was applied in the calculations.

For comparison, calculations were performed using quasi-free kinematics, defined by the conditions $\bvec{p}_j=0, E_j=m_N$, which allows
to factorize the elementary amplitudes out of the momentum integration. This kinematics is supported by the largest absolute value of the DWF.
In other words, in this case the off-shell states of the struck nucleon are ignored, since they are suppressed by the DWF. 

To calculate the factor $\sqrt{m_d/2E_j(m_d-E_j)}$ in the rescattering amplitudes, it is also necessary to specify the value of $E_j$,
which would reasonably be taken to be the same as in the amplitudes of elementary transitions.
However, for simplicity, in the default calculations this factor was replaced by $m_N^{-1/2}$ in accordance with previous
GEA studies \cite{Frankfurt:1996uz,Larionov:2022gvn}, since large momenta of spectator are suppressed by the DWF.
Accordingly, in the IA amplitude of Eq.(\ref{M_IA_tot})
the factors $\sqrt{E_i/(m_d-E_i)},~i=3,4,5$ were replaced by unity and $m_d$ -- by $2m_N$.
The impact of these simplifications is discussed in section \ref{Disc}.

\section{Cross section of backward proton production}
\label{Xsection}

The full differential cross section of the process $pd \to ppn$ is expressed as
\begin{equation}
    d\sigma = \frac{(2\pi)^4\overline{|M|^2}}{4I_{1d}} d\Phi~,      \label{dsigma}
\end{equation}
where $\overline{|M|^2}$ is the modulus squared of the invariant matrix element summed over spin projections of the outgoing particles
and averaged over spin projections of the incoming ones, $I_{1d}=[(p_1p_d)^2-m^2m_d^2]^{1/2}$ is the flux factor, and
\begin{equation}
  d\Phi = \delta^{(4)}(p_1+p_d-p_3-p_4-p_5) \frac{d^3p_3}{(2\pi)^32E_3} \frac{d^3p_4}{(2\pi)^32E_4} \frac{d^3p_5}{(2\pi)^32E_5}    \label{dPhi}
\end{equation}
is the invariant three-body phase space volume element.

The matrix element in Eq.(\ref{dsigma}) was calculated using the partial amplitudes of Figs.~\ref{fig:IA},\ref{fig:el} and \ref{fig:inel}.
Thus, the full calculation is based on the expression
\begin{equation}
   M = M_{\rm IA} + M_{\rm el} + M_{\rm inel}~,       \label{Mfull}
\end{equation}
where the IA term is given by Eq.(\ref{M_IA_tot}), the elastic term is expressed by Eq.(\ref{M_el_fin}),
and the inelastic one -- by Eq.(\ref{M_inel_fin}).
To save CPU time, partial amplitudes with CEX were neglected in $M_{\rm IA}$ and $M_{\rm el}$ in default calculations.
Their impact is quite moderate and is discussed in sec. \ref{Disc}.

The partial differential cross section of the backward proton (proton 4 in our notation) production was calculated as
\begin{equation}
  E\frac{d\sigma}{d^3p} = E \frac{(2\pi)^4}{4I_{1d}} \Phi \langle \overline{|M|^2} \delta^{(3)}(\bvec{p}-\bvec{p}_4) \rangle~,     \label{dsigmad3p}
\end{equation}
where $\Phi = \int d\Phi$ is the total integrated three-body phase space volume.
The averaging over phase space volume $\langle \ldots \rangle$ in Eq.(\ref{dsigmad3p}) was calculated by the Monte-Carlo method. 

\begin{figure}
  \begin{center}
     \includegraphics[scale = 0.6]{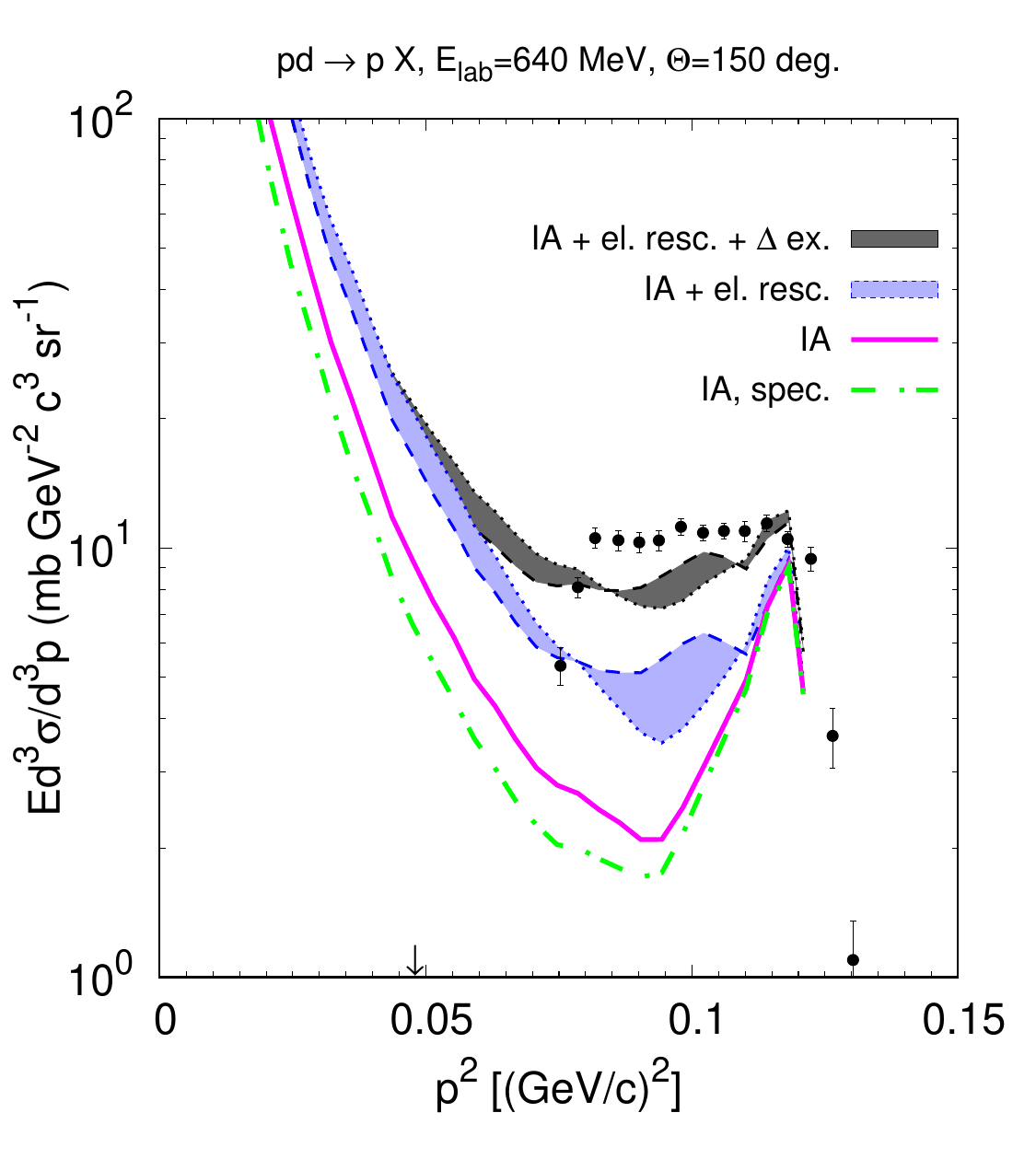}
  \end{center}
  \caption{\label{fig:pd_640mev_p} Differential cross section of proton production in $pd \to ppn$ reaction at the proton beam energy of 640 MeV at the polar angle $\Theta=150\degree$
    vs proton momentum squared.
    Full calculation is shown by the dark (gray) band while the calculation without intermediate $\Delta$ resonance -- by the light (blue) band.
    The borders of the bands are obtained by using the on-shell intermediate state kinematics (dashed line) and the quasi-free kinematics (dotted line).
    The calculation in the IA is shown by the solid (magenta) line. The contribution of the spectator amplitude of Fig.~\ref{fig:IA}b is shown by the dash-dotted
    (green) line.
    Experimental data for inclusive reaction $pd \to p X$ are from Ref.~\cite{Ero:1981jj}. Vertical arrow shows the maximum momentum squared, 0.048 (GeV/c)$^2$,
    up to which the one-pion production in the final state is possible.}
\end{figure}

Fig.~\ref{fig:pd_640mev_p} shows the spectrum of protons at the polar angle $\Theta=150\degree$ from the $pd \to ppn$ process at a proton beam energy of 640 MeV.
It is interesting that this energy almost exactly corresponds to the $NN \to N\Delta$ reaction threshold on nucleon at rest assuming that the $\Delta$ is produced on the mass shell. 
The IA calculation completely fails to describe experimental data, except the region near the kinematic limit where the spectator amplitude of Fig.~\ref{fig:IA}b
dominates.  This can be understood in the following way.
With increasing momentum of the backward proton 4 the invariant energy 
$\sqrt{s_{12}}=\sqrt{(p_1+p_2)^2}$ decreases and, therefore, the cross section $\sigma_{pn}$ increases (see Fig.~\ref{fig:sigtot&alpha}).
As a result, the spectator amplitude becomes leading, pushing the cross section upward, in agreement with Ref.~\cite{Furui:1978tf}.
Taking into account elastic intermediate states increases the cross-section by 2-3 times.
The intermediate $\Delta$ resonance gives another $50\%$ increase in the cross section.
Calculations with on-shell intermediate states result in a shallower minimum of the cross section
as compared to the calculation with quasi-free kinematics and are in a better agreement with data.

\begin{figure}
  \begin{center}
     \includegraphics[scale = 0.6]{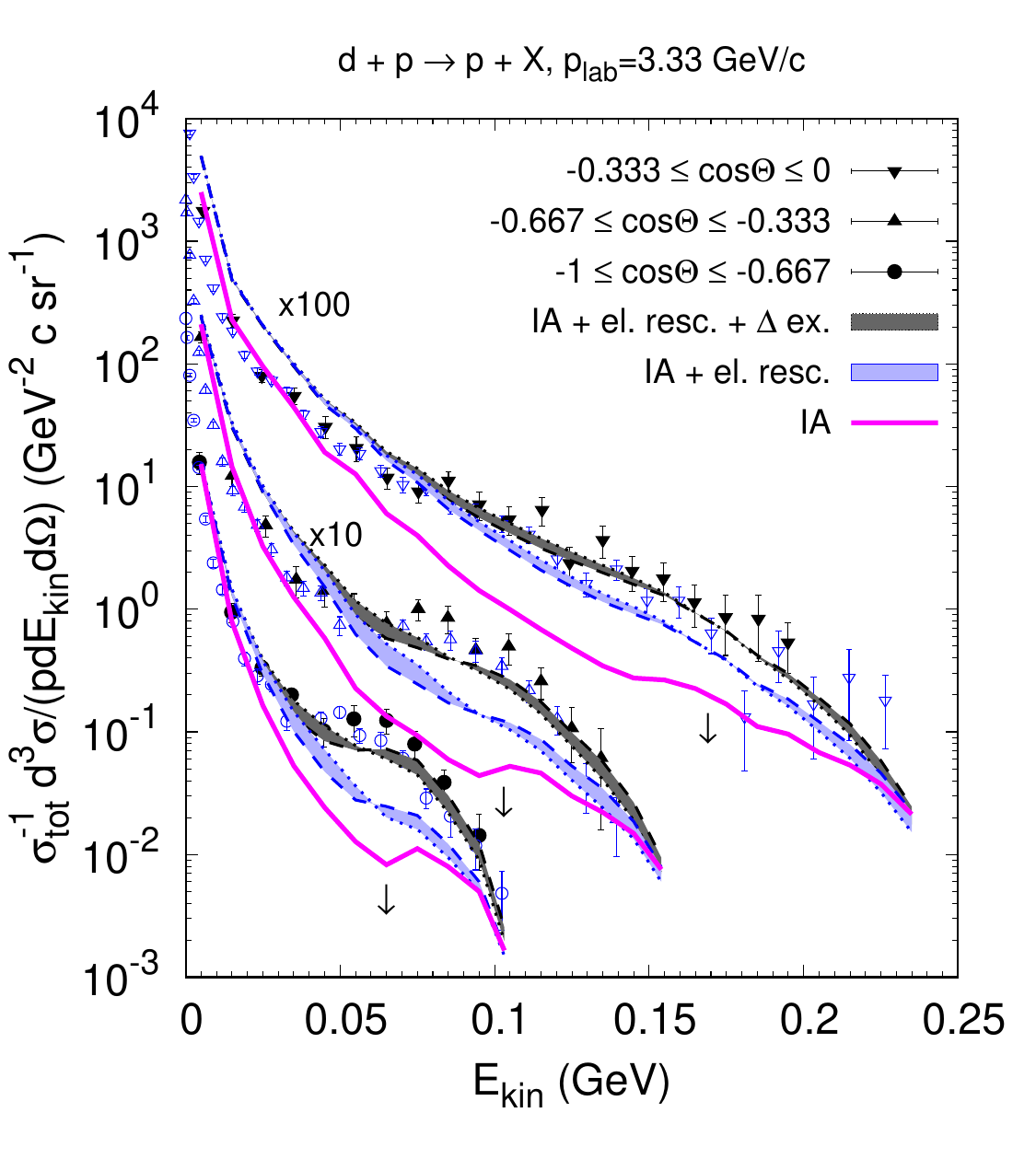}
  \end{center}
  \caption{\label{fig:dp_3.33gevc_p} Kinetic energy differential cross section of proton production in $dp$ collisions at the deuteron momentum
    3.33 GeV/c for $\cos\Theta$ in the intervals $[-0.333:0]$ (upper curves), $[-0.667:-0.333]$ (middle curves), and $[-1:-0.667]$ (lower curves)
    where $\Theta$ is the polar angle of proton momentum relative to the incident proton momentum in the deuteron r.f..
    The calculated results are denoted as in Fig.~\ref{fig:pd_640mev_p}.
    The cross sections are divided by the total $pd$ cross section of 82.889 mb at the proton momentum 1.66 GeV/c \cite{Bugg:1966zz}
    and multiplied by the scaling factors shown on the plot. 
    Experimental data for inclusive reaction $d + p \to p + X$ are from Ref.~\cite{Warsaw-Dubna:1977qek} (filled symbols) and Ref.~\cite{Dubna-Kosice-Moscow-Tbilisi-Warsaw:1997ton}
    (open symbols). The arrows mark the maximum kinetic energies of the proton
    for the final state with one pion production: 65 MeV, 103 MeV, and 169 MeV in the order starting from the most backward polar angle interval.}
\end{figure}

It is natural to expect that the contribution of the intermediate $\Delta$ resonance should increase with increasing proton beam energy.
Fig.~\ref{fig:dp_3.33gevc_p} shows the differential cross section of proton production in the $d p \to p p n$ process at the deuteron momentum of 3.33 GeV/c
which corresponds to the proton beam energy 0.97 GeV.
The proton kinetic energy spectra are shown in three polar angle ranges
of the backward hemisphere in the r.f. of the deuteron.
The IA calculation agrees with data at small kinetic energy, $E_{\rm kin}$,  but strongly underestimates the cross section
at large $E_{\rm kin}$.
Including elastic rescattering amplitudes increases the cross section by two-three times at large $E_{\rm kin}$ and, in the range of smallest polar angles, 
also at small $E_{\rm kin}$.
The latter effect can be regarded as spurious, as the GEA method is inapplicable for too slow particles.
\footnote{This limitation arises from neglecting the Fermi motion in the deuteron when calculating the parameters $\Delta_l$,
see Eqs.(\ref{Delta_l}),(\ref{Delta_l_Delta}). This is legitimate when momenta of all outgoing particles are much larger than
typical momenta in the deuteron $\sim \sqrt{\langle p^2 \rangle} =0.13$ GeV/c for the DWF in the Paris model.}

Further increase of the cross section is obtained when including the $\Delta$ resonance as
an intermediate state. The effect of the $\Delta$ resonance is especially pronounced at the largest polar angles where it reaches a factor of three
and substantially improves agreement with experiment.
However, the experimental cross sections still remain somewhat underpredicted.   

\begin{figure}
  \begin{center}
     \includegraphics[scale = 0.6]{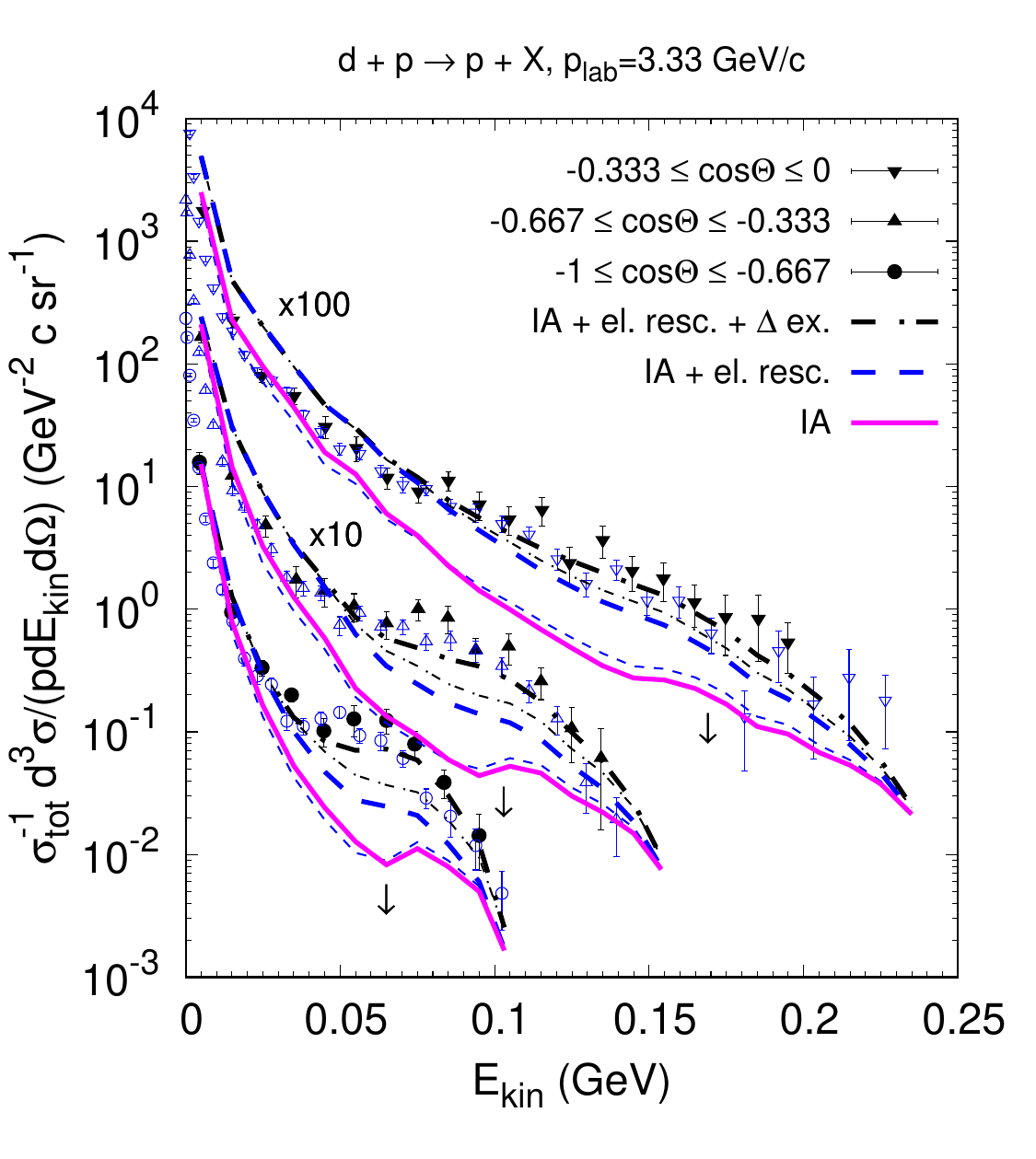}
  \end{center}
  \caption{\label{fig:dp_3.33gevc_p_expl} Same as in Fig.~\ref{fig:dp_3.33gevc_p} but for rescattering amplitudes with and without
    slow particles in the intermediate rescattering state.
    Dash-dotted (black) lines -- calculation including IA, all elastic rescattering amplitudes, and amplitudes with $\Delta$ resonance
    in the intermediate state. Thick lines -- all amplitudes with intermediate $\Delta$ (full calculation). Thin lines -- same but without amplitudes (a) of Fig.~\ref{fig:inel}.
    Dashed (blue) lines -- calculation including IA and elastic rescattering amplitudes. Thick lines -- all elastic rescattering amplitudes. Thin lines -- same but without amplitudes (a)
    of Fig.~\ref{fig:el}.
    Solid (magenta) lines -- calculation in the IA.}
\end{figure}

To see which diagrams are responsible for the enhancement of backward production, Fig.~\ref{fig:dp_3.33gevc_p_expl} shows the calculations without including
slow particles in the intermediate rescattering states which are described by the diagrams (a) in Figs.~\ref{fig:el},\ref{fig:inel}.
Taking into account only the fast intermediate states of nucleons gives results very close to the IA.
Moreover, at small $E_{\rm kin}$, as expected, this leads to the typical Glauber absorption effect due to the interference
of the IA and rescattering amplitudes.
Adding fast intermediate $\Delta$ resonance states gives just a slight increase of the cross section relative to the calculation with all intermediate nucleon states. 
Thus, the enhancement of backward production is mostly due to the slow nucleons and $\Delta$ resonances in the intermediate rescattering states
which is a consequence of the factor $|\bvec{p}_l|^{-1}$ in Eqs.(\ref{M_el_fin}),(\ref{M_inel_fin}) that originates from the eikonal representation of the propagators,
Eqs.(\ref{lprimeInvProp}),(\ref{lprimeInvProp_Delta}).

\begin{figure}
  \begin{center}
     \includegraphics[scale = 0.6]{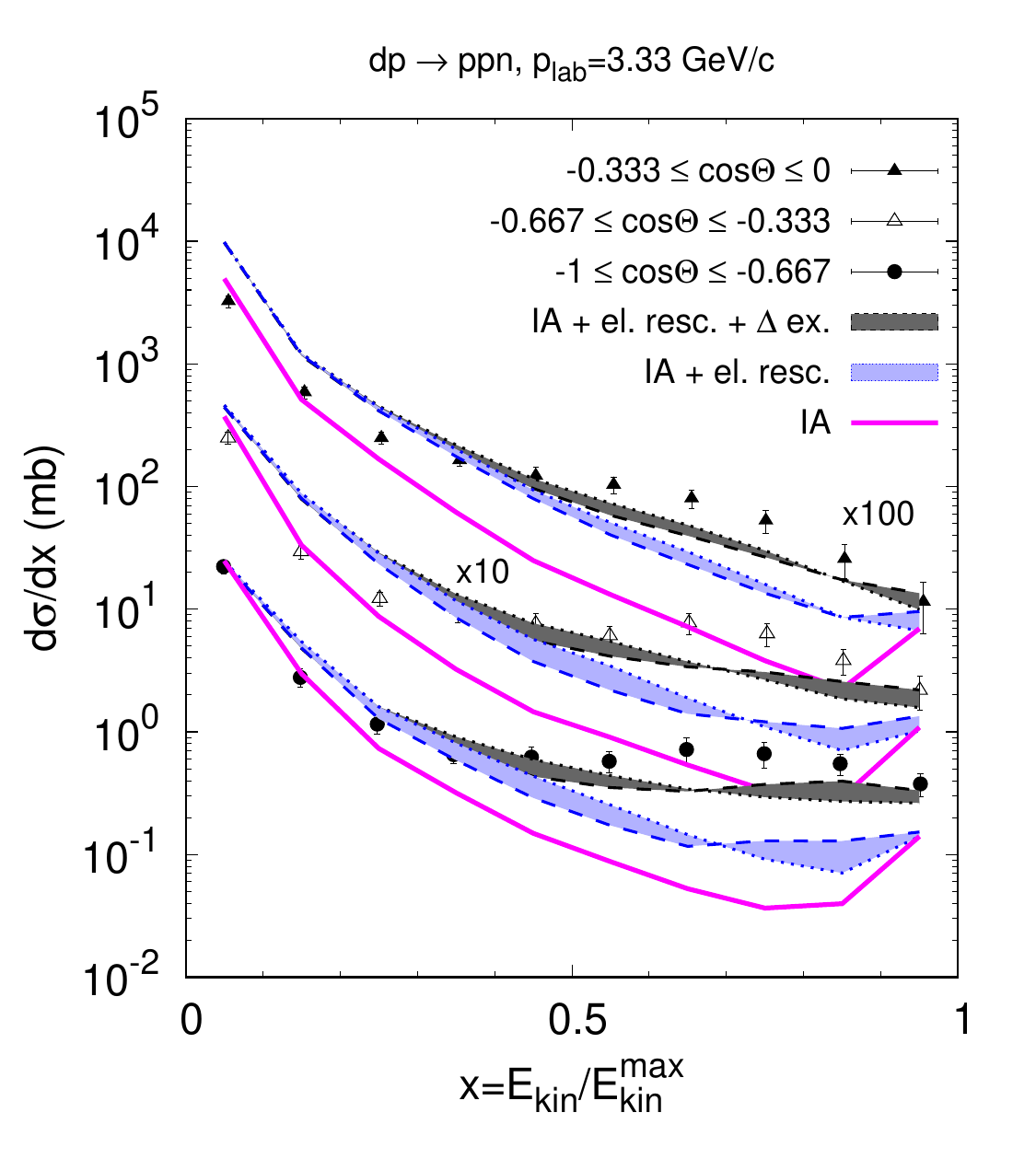}
  \end{center}
  \caption{\label{fig:dp_3.33gevc_x_p} Differential cross section of proton production in $dp$ collisions at the deuteron beam momentum 3.33 GeV/c as a function of the ratio of the proton kinetic energy
    to the maximum possible proton kinetic energy at the given polar angle in the deuteron rest frame. Calculated results are denoted in the same way as in Fig.~\ref{fig:pd_640mev_p}.
    Experimental data for exclusive reaction channel $d p \to p p n$ are from Ref.~\cite{Warsaw-Dubna:1977qek}.}
\end{figure}

Interestingly, the discrepancy with data becomes smaller at the kinematic limit which is given by the proton
scattered with the maximum possible kinetic energy, i.e. when the relative momentum of the recoiling (forward moving) proton and neutron tends to zero.
This kinematic regime is better visible in Fig.~\ref{fig:dp_3.33gevc_x_p} which shows the distribution of the backward protons in variable
$x$ defined as the ratio of the proton kinetic energy to the maximum possible at the given polar angle kinetic energy $E_{\rm kin}^{\rm max}$.
The latter is determined by the condition that the relative momentum of the recoiling $pn$ system is zero.
The differential cross section calculated in the IA grows with increasing $x$ near $x=1$ which is explained by increasing $pn$ total cross section with decreasing invariant energy $\sqrt{s_{12}}$
(see discussion of Fig.~\ref{fig:pd_640mev_p} above).
Note that the growth is not present in the energy spectra of Fig.~\ref{fig:dp_3.33gevc_p} since the kinematic limit $E_{\rm kin}^{\rm max}$ depends on the polar angle and is therefore blurred 
by wide angular ranges. 

\begin{figure}
  \begin{center}
     \includegraphics[scale = 0.6]{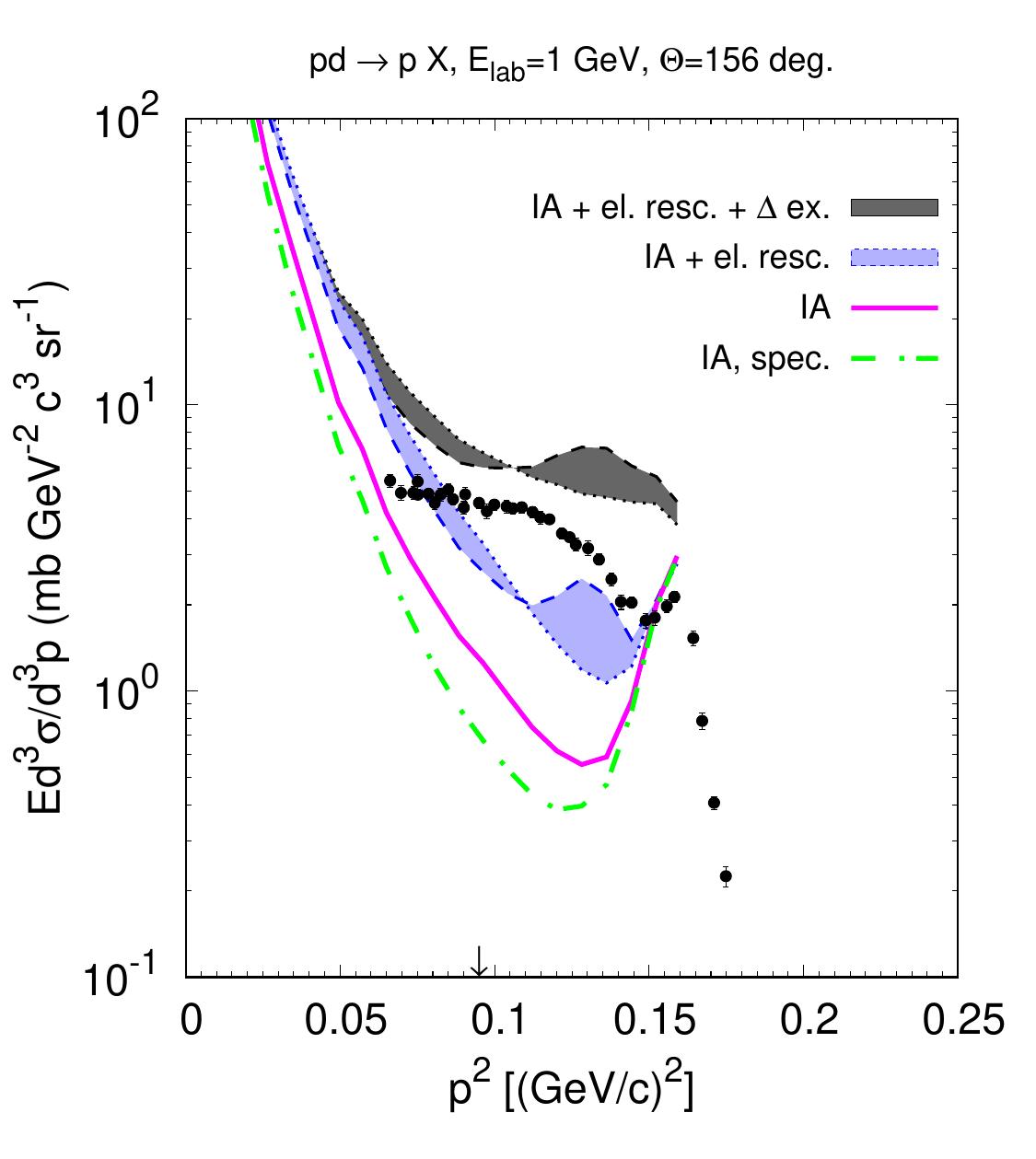}
  \end{center}
  \caption{\label{fig:pd_1gev_p} Same as in Fig.~\ref{fig:pd_640mev_p} but for the proton beam energy of 1 GeV and the polar angle $\Theta=156\degree$.
    Experimental data are from Ref.~\cite{Andronenko:1983eqn}. Vertical arrow shows the limit, 0.095 (GeV/c)$^2$,
    up to which the one-pion production in the final state is possible.}
\end{figure}

Fig.~\ref{fig:pd_1gev_p} shows the differential cross section of proton production at the proton beam energy 1 GeV in comparison with high resolution data of Ref.~\cite{Andronenko:1983eqn}.
In agreement with results at 640 MeV, the IA alone is already enough to describe the experimental cross section close to the maximum proton momentum squared.
However, the IA calculation strongly underestimates the data at
smaller $p^2$, in particular, above 0.095 (GeV/c)$^2$ where only the exclusive $pd \to ppn$ channel is open.
Taking into account elastic intermediate states significantly increases the cross section, but is still insufficient to achieve experimental values.
Including the intermediate $\Delta$ resonance leads to a further strong increase in the cross sections.
However, this results in too flat behavior of the cross section as a function of $p^2$, which leads to an overestimation of the data near the kinematic limit.
Most likely, the reason for the overestimation lies in missing final state interaction between recoiling proton and neutron, which have a small relative momentum near the kinematic limit.

\begin{figure}
  \begin{center}
      \includegraphics[scale = 0.6]{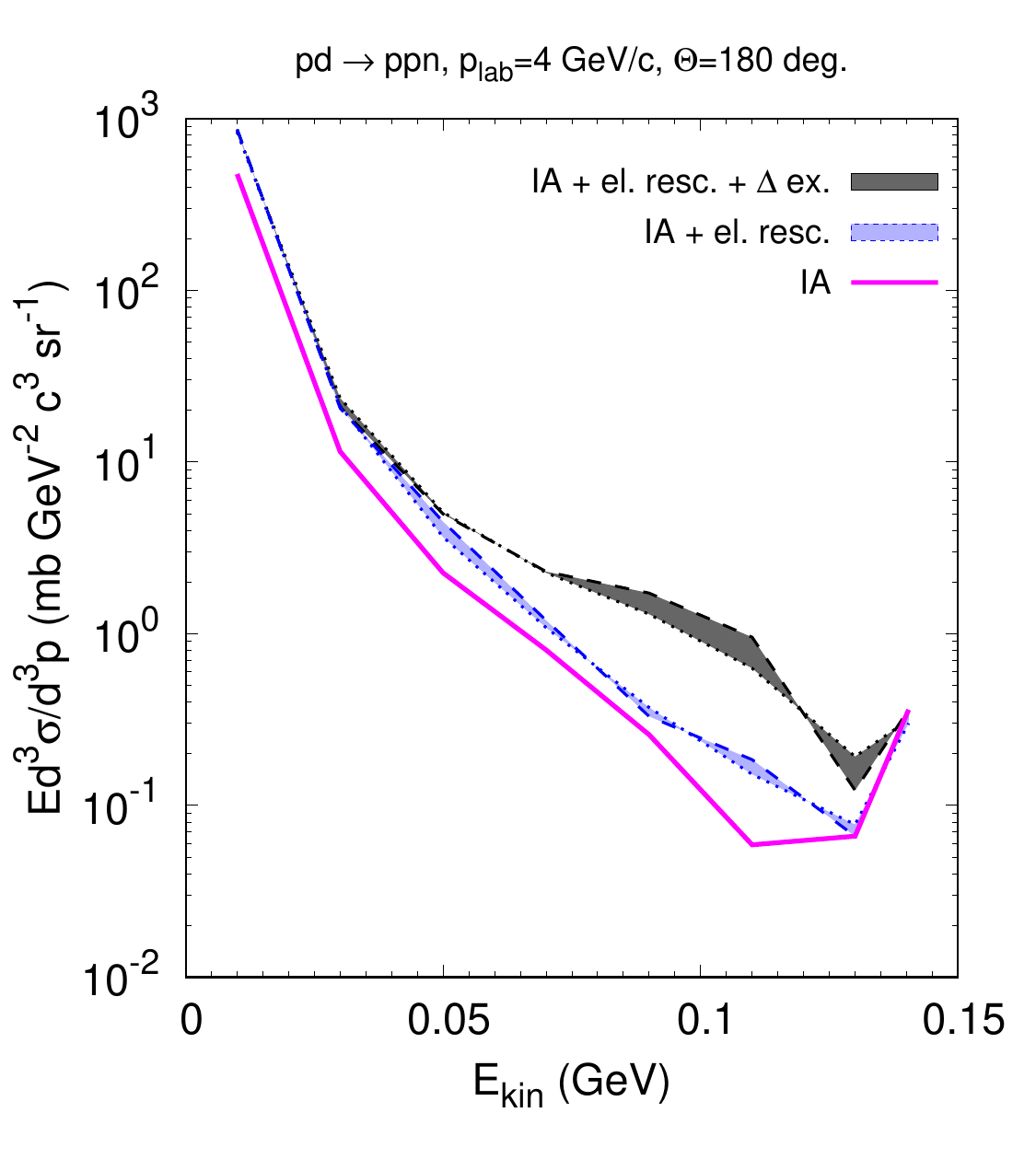}
  \end{center}
    \caption{\label{fig:pd_4gevc_p} Differential cross section of proton production in $pd \to ppn$ reaction at the proton beam momentum of 4 GeV/c
    at the polar angle $\Theta=180\degree$
    vs proton kinetic energy. Notation of calculated results is the same as in Fig.~\ref{fig:pd_640mev_p}.}
\end{figure}

\begin{figure}
  \begin{center}
      \includegraphics[scale = 0.6]{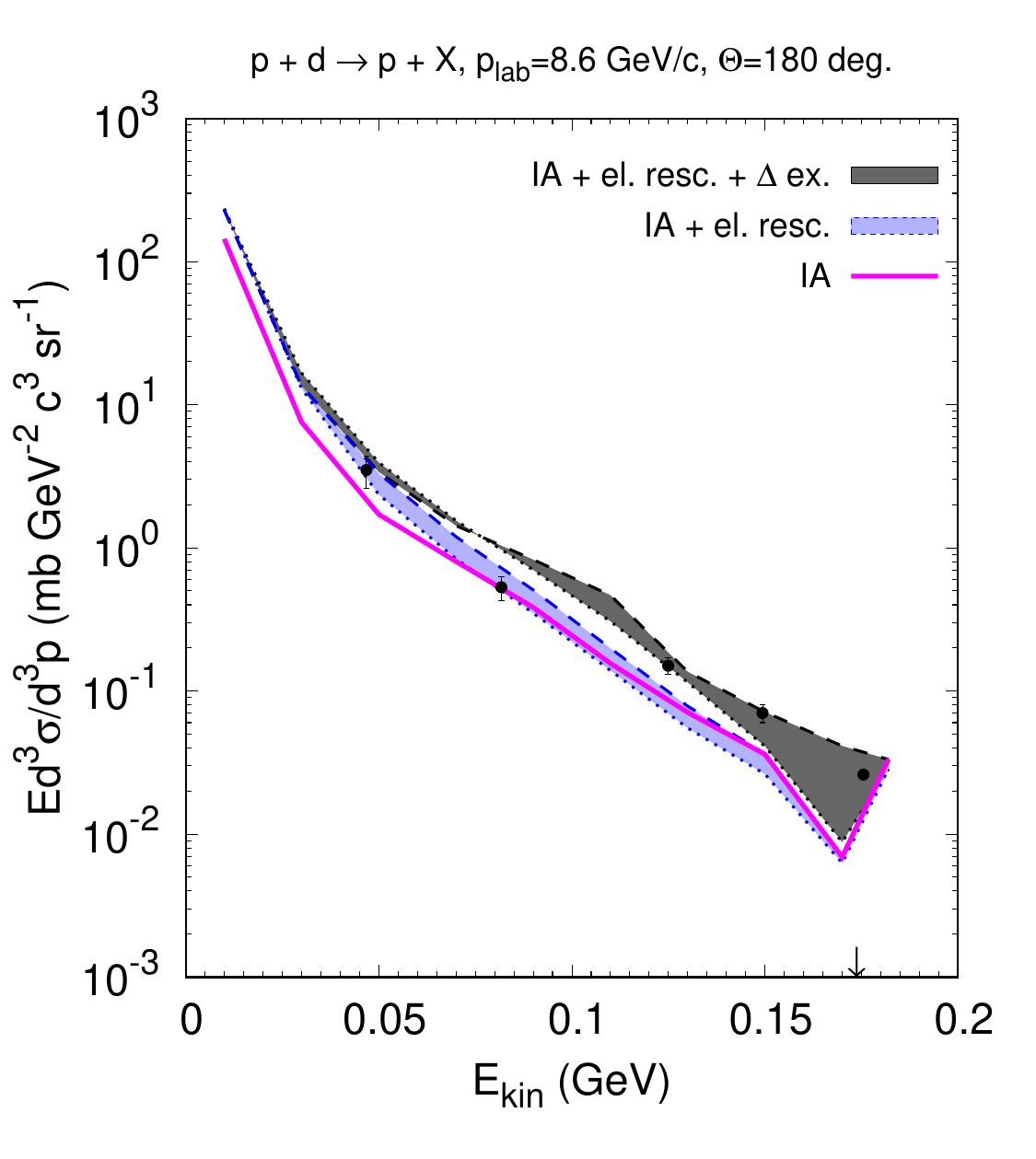}
  \end{center}
    \caption{\label{fig:pd_8.6gevc_p} Same as in Fig.~\ref{fig:pd_4gevc_p} but for the proton beam momentum of 8.6 GeV/c  
    Experimental data for inclusive reaction $pd \to p X$ at 8.6 GeV/c are from Ref.~\cite{Baldin:1977iu}. Vertical arrow shows the maximum kinetic energy, 0.174 GeV,
    up to which the one-pion production in the final state is possible.}
\end{figure}

\begin{figure}
  \begin{center}
      \includegraphics[scale = 0.6]{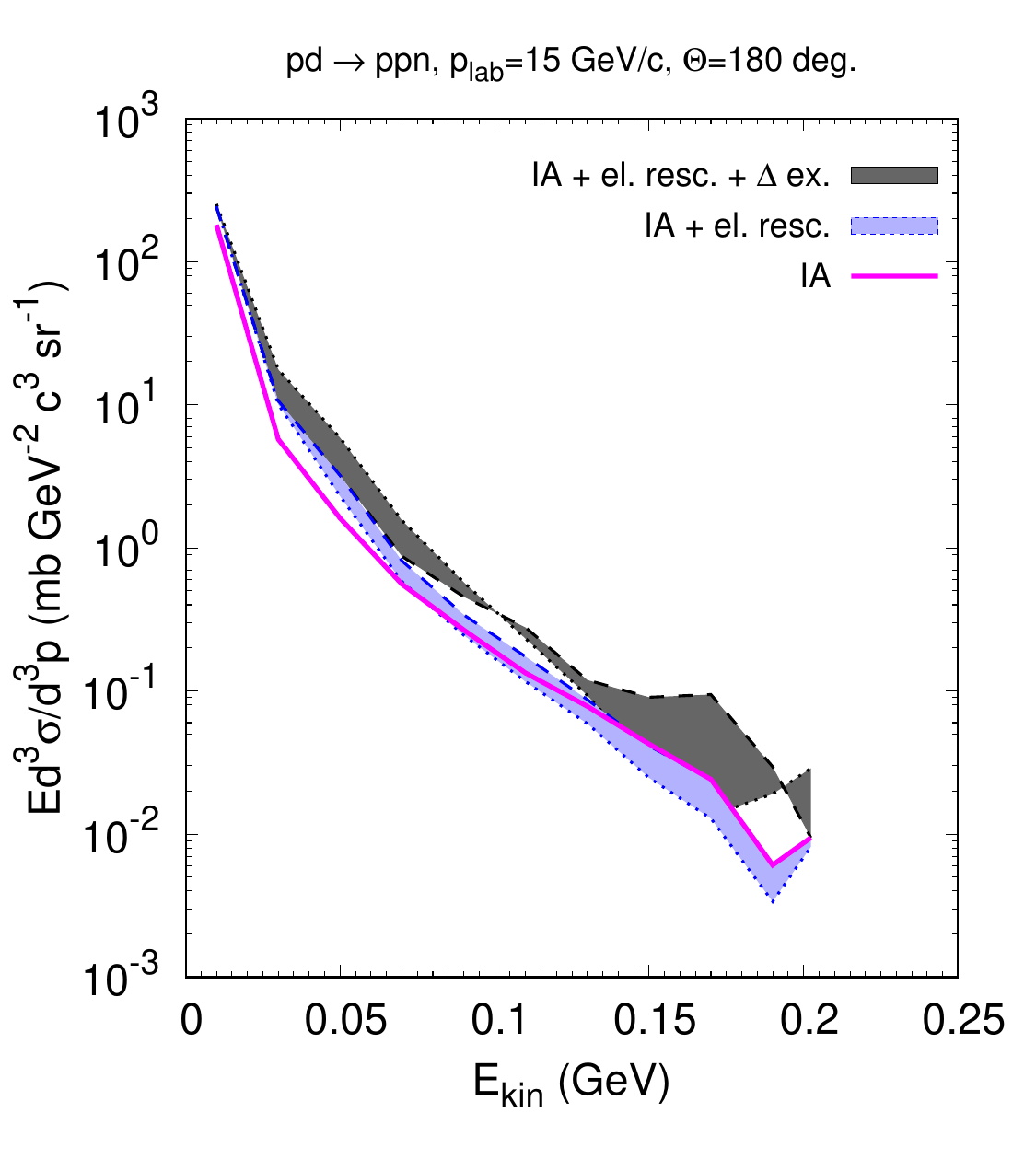}
  \end{center}
    \caption{\label{fig:pd_15gevc_p} Same as in Fig.~\ref{fig:pd_4gevc_p} but for the proton beam momentum of 15 GeV/c.}
\end{figure}

Figs.~\ref{fig:pd_4gevc_p},\ref{fig:pd_8.6gevc_p}, and \ref{fig:pd_15gevc_p} show the energy differential cross section of proton production at $\Theta=180\degree$
for the proton beam momenta of 4, 8.6, and 15 GeV/c, respectively. This beam momentum range will be accessible, for example, at the NICA SPD facility
\cite{Abramov:2021vtu}.
The calculation in the IA produces kinks in the kinetic energy spectrum at 0.11, 0.17, and 0.19 GeV  for $p_{\rm lab}=4$, 8.6, and 15 GeV/c, respectively.
These kinks correspond to the minimum of the total $pn$ cross section at $\sqrt{s_{12}}=2.07$ GeV ($p_{\rm lab}=0.97$ GeV/c, see Fig.~\ref{fig:sigtot&alpha}) in vicinity of
the pion production threshold.
The sensitivity to the elastic and inelastic intermediate state is highest near the kinks.
The kinks are explained as follows.
The actual shape of the $E_{\rm kin}$ spectrum in the IA is governed by the spectator graph of Fig.~\ref{fig:IA}b and is the result of folding the momentum dependence of the DWF squared
with $\sqrt{s_{12}}$ dependence of the total $pn$ cross section.
The kinks correspond to the spectator momenta of 0.48, 0.59, and 0.64 GeV/c, where the DWF is dominated by the $D$ wave.
Thus, the slope of the momentum dependence of the DWF is relatively small
and the shape of the $E_{\rm kin}$ spectrum reflects the shape of $\sqrt{s_{12}}$ dependence of
$\sigma_{pn}$, of course, with account for the bin size of the spectrum needed to minimize statistical error of the Monte-Carlo procedure ($\sim 15\%$).  
At lower beam momenta, the deep and quite broad valleys in $E_{\rm kin}$ (Fig.~\ref{fig:dp_3.33gevc_p})
and $p^2$ spectra (Figs.~\ref{fig:pd_640mev_p},\ref{fig:pd_1gev_p}) are present instead of kinks since the DWF is dominated by the $S$ wave having a steeper slope of the momentum dependence.

At  $p_{\rm lab}=8.6$ GeV/c, the inclusion of the intermediate $\Delta$ resonance rises the cross section at high values of $E_{\rm kin}$ to a good agreement with data. 
It should be emphasized that this rise occurs in the case of elementary amplitudes with intermediate states on the mass shell, and not in the case of quasi-free amplitudes.
In contrast, the calculations with quasi-free amplitudes do not much change the cross sections at high $E_{\rm kin}$ relative to the IA case.

Similar trends persist also at $p_{\rm lab}=15$ GeV/c except the very end of the spectrum where the calculation with quasi-free amplitudes and  intermediate $\Delta$
yields the largest cross section. It is also evident that the sensitivity to the choice of the kinematics of the elementary amplitudes with intermediate $\Delta$ resonance becomes
stronger with increasing beam momentum. This is mostly a consequence of the sharp momentum-transfer dependence of the pion propagators in Eq.(\ref{M_klm_1ij}). 


\section{Discussion}
\label{Disc}

Despite the use of the usual non-relativistic DWF, the deuteron vertex factor included in Eq.(\ref{M_IA^a}) 
leads to a relativistically consistent differential cross section on the deuteron:
\begin{equation}
  d\sigma_{1d \to 345} = \frac{v_{12}}{v_1} |\phi(-\bvec{p}_5)|^2 d^3p_5 d\sigma_{12 \to 34}~,  \label{dsigma_1d345}
\end{equation} 
where $v_1=|\bvec{p}_1|/E_1$ is the beam proton velocity, $v_{12}=I_{12}/E_1p_2^0$ is the relative velocity of the protons 1 and 2
defined according to Ref.~\cite{BLP} with $I_{12}=\sqrt{(p_1p_2)^2-m_N^2p_2^2}$ being Lorentz invariant flux factor.
\begin{equation}
   d\sigma_{12 \to 34} = \frac{(2\pi)^4\delta^{(4)}(p_3+p_4-p_1-p_2)}{4I_{12}} |M(3,4;1,2)|^2 \frac{d^3p_3}{(2\pi)^32E_3} \frac{d^3p_4}{(2\pi)^32E_4}  \label{dsigma_1234}
\end{equation}
is the Lorentz invariant differential cross section of the process $1 2 \to 3 4$.
The quantities $v_1$, $v_{12}$ and the product $|\phi(-\bvec{p}_5)|^2 d^3p_5$ are defined in the deuteron r.f. and, therefore, are by definition Lorentz invariant.
Thus, the use of such a vertex factor provides a kind of ``minimal'' relativization, neglecting genuine relativistic DWF effects,
such as the appearance of the $P$-wave and negative energy states \cite{Gilman:2001yh,Sargsian:2022rmq}.

\begin{figure}
  \begin{center}
     \includegraphics[scale = 0.6]{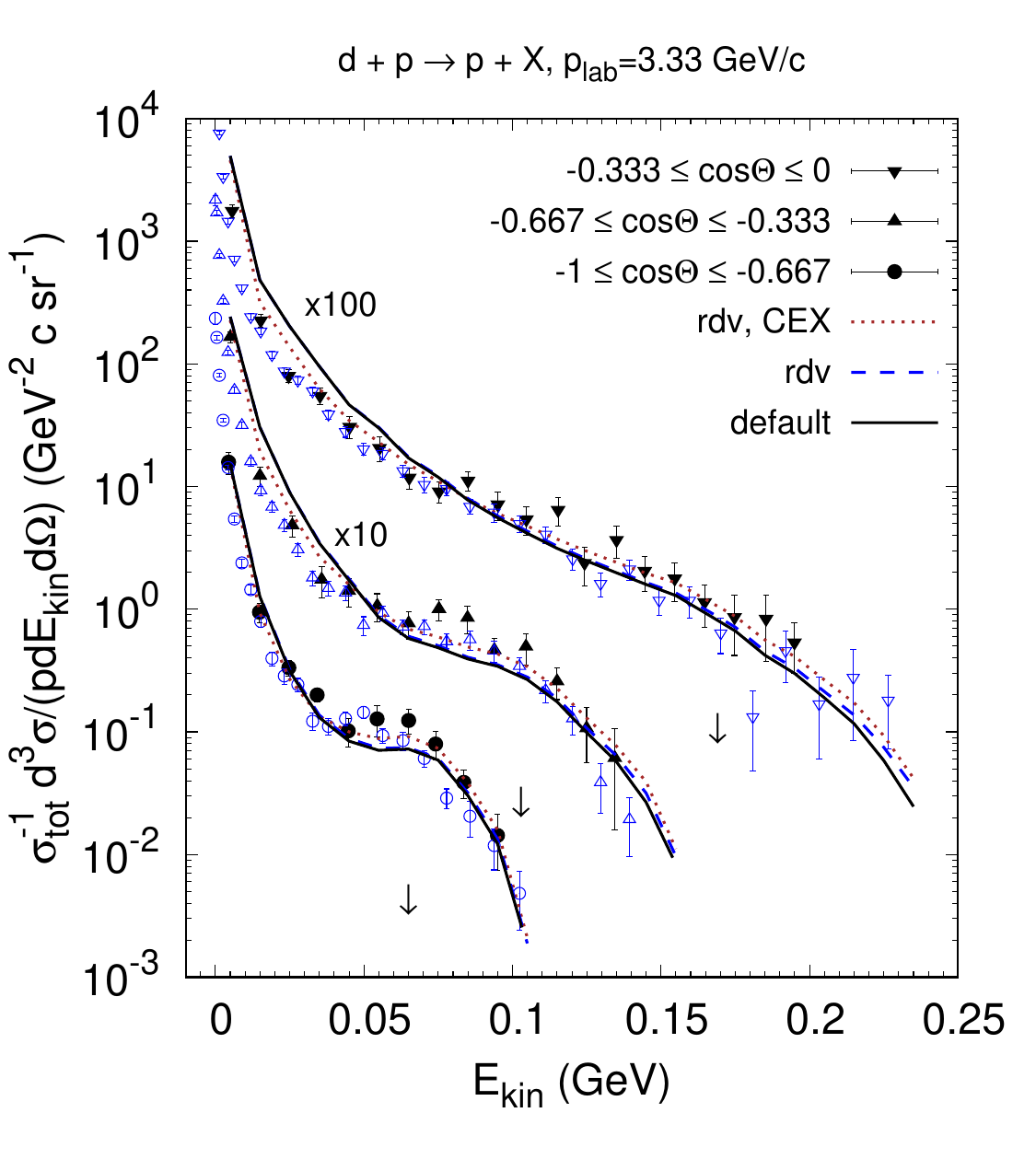}
  \end{center}
  \caption{\label{fig:dp_3.33gevc_p_cex} Same as in Fig.~\ref{fig:dp_3.33gevc_p} but for calculations including the correct deuteron vertex factors (dashed lines)
    and both correct vertex factors and CEX diagrams (dotted lines) in comparison with default calculation with on-shell intermediate state kinematics (solid lines).
    All calculations include IA amplitudes, elastic rescattering amplitudes, and amplitudes with $\Delta$ resonance in the intermediate state.}
\end{figure}

One of the approximations used in obtaining all previous numerical results was to replace the proton and neutron energies in the deuteron vertex factor
with the nucleon mass, which would lead to an additional factor $p_2^0/E_5$ on the r.h.s. of Eq.(\ref{dsigma_1d345}), thereby violating the conservation of baryon number
in the process $1 d \to 3 4 5$.
It is instructive to check how great the influence of the choice of the vertex factor is.
Fig.~\ref{fig:dp_3.33gevc_p_cex} compares the calculation with the correct vertex factors in all diagrams (dashed lines) with the default calculation (solid lines)
for case of $d p \to p p n$ reaction at the deuteron momentum of 3.33 GeV/c. 
It can be seen that the vertex factors have virtually no effect on the resulting energy spectra, except for a slight enhancement at the kinematic boundary.
The enhancement is bigger in pure IA calculations (not shown), but it is masked by rescattering contributions.

Another approximation was to neglect the CEX diagrams, see Eq.(\ref{M4312_CEX}). Since they are governed by the $\pi$ and $\rho$ exchange, their contribution
is expected to be larger at lower beam energies. As shown in Fig.~\ref{fig:dp_3.33gevc_p_cex} (dotted lines), for the reaction $d p \to p p n$ at 3.33 GeV/c
the CEX diagrams increase the proton production cross section at high energies by approximately 30\% and also suppress the cross section at low energies, improving
the agreement with experiment.
 
This analysis does not take into account multiple rescattering diagrams (three-step or more) \cite{Frankfurt:1996uz,Larionov:2024ynh}.
They are expected to influence the spectra at high momenta of the produced particle.
Although it seems more efficient to calculate such terms in coordinate space,
the difficulty lies in the non-factorizability of the soft rescattering amplitudes.
It should be recalled that in all previous studies using the GEA method, one of the amplitudes was hard and for this reason was factorized out of the momentum integrals.
Of course, such calculations will lead to many new diagrams that will require more computing resources, but which can be considered feasible in the near future.
\footnote{The full calculation with CEX (dotted lines in Fig.~\ref{fig:dp_3.33gevc_p_cex})
required about 72 hours of CPU time on ten 3.70GHz processors.}

As mentioned in sec.~\ref{intro}, most of the data on cumulative proton production are inclusive and, therefore pion production channels should be taken into account.
The main mechanism is due to the formation and decay of the intermediate $\Delta(1232)$ resonance.
At the proton beam momenta of about 1.3-3 GeV/c, the sum of
$pn \to p\Delta^0$ and $pn \to n\Delta^+$ cross sections (each of them is 1/3 of the $pp \to n\Delta^{++}$ cross section) is comparable with elastic $pn$ cross section.
Thus, the channels $pd \to pNN\pi$ are expected to slightly increase the low-energy part of the cumulative proton spectrum, however, are suppressed by phase space restrictions
at its high energy part.
The evaluation of the pion production channels was beyond the scope of this work, but certainly deserves to be carried out in future theoretical studies.

Finally, one more remark should be made regarding the elementary $NN$ amplitudes.
The high-energy parameterizations of Eqs.(\ref{M3412}),(\ref{M4312_CEX})
are applicable at $p_{\rm lab} \gtsim 1$ GeV/c (or $\sqrt{s} \gtsim 2.1$ GeV).
Although this seems sufficient at first glance, the cumulative proton spectra in the high-energy range
are determined by the spectator amplitude of Fig.~\ref{fig:IA}b and, therefore, depend on the off-shell $pn$ amplitude at $\sqrt{s} \simeq 2m_N$,
for which Eqs.(\ref{M3412}),(\ref{M4312_CEX}) were still used. 
Thus, the use of more sophisticated low-energy $NN$ amplitudes may lead to a change in the obtained results near the high-energy end
of the cumulative proton spectrum.

\section{Summary and conclusions}
\label{summary}

The proton production in the exclusive $p d \to p p n$ process in the backward hemisphere in the deuteron rest frame
was studied on the base of Feynman diagrams. The calculations took into account the one-step IA amplitudes with elastic scattering of the incoming
proton on the proton and neutron as well as the two-step rescattering amplitudes with nucleon (elastic) and $\Delta$ resonance (inelastic) intermediate states.
The $\Delta$ resonance production/absorption amplitudes, $NN \leftrightarrow N\Delta$, were calculated in the pion exchange model.
The antisymmetry of the full reaction amplitude with respect to the interchange of the outgoing nucleons as well as the isoscalar character of the DWF
were taken into account. The propagator of the intermediate scattering state was expressed in the eikonal form using the GEA,
which allowed integration over the longitudinal component of the momentum of the intermediate spectator nucleon.
Thereby, the pole structure of the propagator was removed which substantially simplifies numerical calculations
of the two-step amplitudes.
Although this goes beyond the simplest Glauber approximation, i.e. the real part of the propagator was taken into account,
the constraint that the emitted particles should not be too slow still remains.
Hence, this approach is applicable
when the momentum of the slowest outgoing nucleon significantly exceeds the relative momentum of the proton and neutron in the deuteron $\sqrt{\langle p^2 \rangle} = 0.13$ GeV/c.
According to this limitation, it is possible to describe the region of kinetic energies above 50 MeV in the spectra of backward protons.

The simplest calculations in the IA greatly underestimate the experimental cross sections.
Elastic intermediate states increase cross sections by a factor of two-three
at the proton beam momenta $p_{\rm lab} \ltsim 4$ GeV/c, but their contribution becomes small at higher beam momenta.
$\Delta$ resonance intermediate states provide a further increase by a factor of two-five.
Their maximum effect is reached at $p_{\rm lab} \sim 4$ GeV/c.
It was found that the increase is mainly caused by partial amplitudes with slow intermediate $\Delta$ resonance states (Fig.~\ref{fig:inel}a).

The results of calculations with full inclusion of the elastic and $\Delta$ resonance intermediate states are in a reasonable agreement with
the data on backward proton production in $pd \to pX$ reaction from Dubna:
\cite{Ero:1981jj} at $p_{\rm lab}=1.27$ GeV/c ($E_{\rm lab}=640$ MeV), $\Theta=150\degree$
and \cite{Baldin:1977iu} at $p_{\rm lab}=8.6$ GeV/c, $\Theta=180\degree$.
At the same time, the calculation substantially underestimates the Dubna data \cite{Warsaw-Dubna:1977qek} on the backward proton production at $p_{\rm lab}=1.67$ GeV/c
(deuteron beam momentum 3.33 GeV/c) for $\Theta=109-131\degree$ and $\Theta=131-180\degree$ and overestimates the Gatchina data \cite{Andronenko:1983eqn} at $p_{\rm lab}=1.70$ GeV/c
($E_{\rm lab}=1$ GeV) for $\Theta=156\degree$.  

The fact that the angle at which the Gatchina data \cite{Andronenko:1983eqn} were taken falls in the middle of the most backward angular range of the
Dubna data \cite{Warsaw-Dubna:1977qek} creates quite a puzzling situation which was never pointed out in previous theoretical studies, partly because of improper choice of the
angular ranges in calculations. For example, the authors of Ref.~\cite{Haneishi:1985ce} (see their Fig.~5b) reproduce the Dubna data \cite{Warsaw-Dubna:1977qek} with calculation
at fixed $\Theta=140\degree$ which is an extremely rough approximation while the authors of Ref.~\cite{Dakhno:1988qx} (see their Fig.~11) strongly underestimate the both data sets,
\cite{Andronenko:1983eqn} and \cite{Warsaw-Dubna:1977qek}.

The present calculations, like the previous calculations in Refs.~\cite{Haneishi:1985ce,Dakhno:1988qx}, provide the baseline results obtained without introducing any non-nucleon
components of the DWF. Hence, a systematic disagreement with the data may provide additional information on the short-range behavior of the DWF and even point to the presence of the $6q$
or $\Delta\Delta$ configurations in the deuteron.

The opportunities to study backward production in proton interactions with different nuclei already exist at the JINR nuclotron, SIS18@GSI , and J-PARC.
In the near future, the SIS100@FAIR, NICA, and HIAF accelerators may potentially provide much more information on particle production in the cumulative region.  

\bibliography{cumul}

\end{document}